\documentclass{article}

\usepackage[letterpaper,margin=4cm]{geometry}
\usepackage{graphicx}
\usepackage{newtxtext}
\usepackage{newtxmath}
\usepackage{natbib}
\usepackage{soul}
\usepackage{hyperref}
\usepackage{subcaption}
\hypersetup{
    colorlinks = true,
    urlcolor   = blue,
    citecolor  = black,
}

\newcommand{\RomanNumeralCaps}[1]
\linenumbers
\newcommand{\bhat}[1]{\mathbf{\hat{#1}}}
\DeclareMathOperator*{\argmax}{arg\ max}
\usepackage{authblk}

\title{Momentum analysis of complex time-periodic flows}
\author[1]{Benjamin R.S. Freeman}
\author[2]{Robert J. Martinuzzi}
\author[1]{Arman Hemmati}

\affil[1]{Department of Mechanical Engineering, University of Alberta}
\affil[2]{Department of Mechanical and Manufacturing Engineering, University of Calgary}

\begin{document}

\maketitle
\begin{abstract}
    Several methods have been proposed to characterize the complex interactions in turbulent wakes, especially for flows with strong cyclic dynamics. This paper introduces the concept of Fourier-Averaged Navier-Stokes (FANS) equations as a framework to obtain direct insights into the dynamics of complex coherent wake interactions. The method simplifies the interpretations of flow physics by identifying terms contributing to momentum transport at different timescales. The method also allows for direct interpretation of non-linear interactions of the terms in the Navier-Stokes equations. By analysing well-known cases, the characteristics of FANS are evaluated. Particularly, we focus on physical interpretation of the terms as they relate to the interactions between modes at different timescales. Through comparison with established physics and other methods, FANS is shown to provide insight into the transfer of momentum between modes by extracting information about the contributing pressure, convective, and diffusive forces. FANS provides a simply calculated and easily interpreted set of equations to analyse flow physics by leveraging momentum conservation principles and Fourier analysis. The method is applicable to flows with complex cyclic waveforms, including broadband spectral energy distributions.
\end{abstract}

\section{Introduction}
Turbulent wakes and jets exhibit complicated physics related to flow nonlinearity. As a result, there have been many methods proposed over the past decades to investigate the underlying characteristics of turbulent flows \citep[e.g. as detailed in][]{PICARD-spod,noack_hierarchy,high-order-thesis, chenDMD, sieber-spod,  schmidt-bmd}. These methods have been successful in representing the flow dynamics underlying complex systems, but it remains that the interpretation of these representations in the physical domain is often difficult and non-intuitive. We aim to address this shortcoming by introducing a technique that allows for the direct physical interpretation of the forces that affect the flow at each timescale.

There are several approaches to analyse fluid systems through dimensionality reduction. Two well-established methods are Proper Orthogonal Decomposition \citep[POD,][]{lumley} and Dynamic Mode Decomposition \citep[DMD,][]{chenDMD}, which are linear decompositions of the flow dynamics. In POD, Singular Value Decomposition (SVD) is performed on flow data. SVD optimizes the modes in terms of energy content, so the large-scale dynamics are captured compactly. For instance, \citet{wang-POD-TBL} used POD on the streamwise component of a boundary layer to investigate the properties of wall-attached eddies, arguing that the resulting modes were a strong statistical representation of the eddies. \citet{he-POD-ahmed} use POD to compare the processes of wake asymmetry and switching on two Ahmed body models, using the modes to aid analysis of the two mechanisms. While interactions between modes can be identified, e.g. through phase portraits of the temporal dynamics or establishing a dynamical system that relates them \citep{noack_hierarchy}, the physical interpretation of these interactions remains challenging. Likewise, one limitation of POD is that isolating the highest-energy modes does not always capture important dynamics. Low-energy modes can be highly important to the evolution of a flow as noted by \citet{Rowley-rom-review}. 

In Dynamic Mode Decomposition snapshots of the flow variables are used to identify a sparser set of dynamics \citep{schmid_dmd}. DMD ``fits" an approximate linear system that describes the transitions between a series of flow snapshots. There is a rich mathematical background to DMD through its connection to the Koopman operator, as described for instance in \citet{chenDMD}. There are several of variant methods derived by changing the optimization target or adding a mode ranking process, e.g. in the Optimized DMD of \citet{chenDMD} or Recursive DMD of \citet{noack-recursive}. An early example on the topic is the jet study of \citet{schmid-jet}, where it was shown that DMD could successfully isolate the large-scale structures, frequencies, and forcing response of jet flows from PIV and Schlieren snapshots. \citet{gomez-dmd-pipe} used DMD on a turbulent pipe flow, where they successfully linked DMD results about the energy and frequency content of the turbulence to predictions from resolvent analysis. This investigation showed how DMD results could connect to the fundamental fluid dynamics equations. However, this connection required an external method in the form of resolvent analysis. The study of \citet{jang-oscillatory} on oscillatory flow over a surface-mounted circular cylinder showed that DMD was able to accurately model the bed shear stress with relatively few modes, but required a large number of modes to represent the flow around the cylinder body. This highlights the potential difficulties and limitations regarding sparsity in DMD representations. 

POD and DMD are widely and successfully employed, but gaps remain in their applicability and interpretability. POD loses dynamical information due to averaging, and returns modes that are only coherent in space. It can also result in modes that contain contributions from several timescales, reducing interpretability. Meanwhile, DMD does not have an energy optimization procedure, and will restrict modes to single frequencies, which may not be desirable in flows with broadband frequency content. Two distinct methods have been introduced that address these shortcomings. The first corresponds originally to a formulation of POD used by Lumley and named Spectral Proper Orthogonal Decomposition (SPOD) by \citet{PICARD-spod}, which was then formalized by \citet{towne-spod}. The method investigates the decomposition of the cross-spectral density tensor. From flow data, this can be implemented as a series of SVDs performed on ensembles of Fourier modes. The structures identified by this method are claimed by \citet{towne-spod} to evolve coherently in space and time, unlike those identified by basic POD or DMD. The modes found by this method correspond to individual frequencies due to the Fourier decomposition and are optimized in terms of energy due to the SVD. 

The second method, proposed by \citet{sieber-spod}, achieves spectral separation directly from time-domain data using a filter applied on the correlation matrix, typically the snapshot matrix of \citet{sirovich}. The filter is implemented through a convolution of the correlation matrix coefficients with a windowing function of custom width \citep{sieber-thesis}. The method recovers the Fourier modes and Power Spectral Density of the flow in the limit of filtering over the whole window. This approach, also identified as SPOD by \citet{sieber-spod} and several subsequent studies, was able to allow for ``much better separation of individual fluid dynamic phenomena into single modes" \citep{sieber-spod} than either POD or Fourier modes.

While these data analysis methods are useful for educing structures or dynamics of particular flows by offering simplified representations, they conspicuously do not include a direct connection to the physical processes that are involved. This would be useful when the complexity of the full Navier-Stokes equations should be avoided. For instance, a common application is a Galerkin projection of the governing equations onto the spatial modes. This approach is useful when short-term predictions of the flow dynamics are desirable. However, this does not provide spatial information for characterizing regions where nonlinear interactions are important.

Other nonlinear analysis methods include higher-order spectral methods \citep{high-order-thesis}, which introduce the idea of the ``bispectrum". The bispectrum is used to characterize relationships between Fourier modes of a nonlinear system, specifically by measuring the degree of correlation between triads of frequency or wavenumber components. This makes it natural for analysing systems with quadratic nonlinearity, such as incompressible flows described by the Navier-Stokes equations. These triadic interactions between structures at different wavenumbers have been successfully applied, for example to the turbulent kinetic energy cascade \citep{durbin_pettersson_2011}. However, interactions between modes of different frequencies has been less extensively studied. Recently, the Bispectral Modal Decomposition was proposed by \citet{schmidt-bmd}, which is a promising method to analyse triadic interactions through a specialized reduced order method that maximizes the bicorrelation between frequency or wavenumber components of a flow. However, the method cannot directly relate the magnitude of the nonlinear interactions to the physical transportation processes that govern the flow, and relies on coincidence of local components of Fourier modes.

Here, we introduce a new technique, initially proposed in \citet{tsfp-fans}, in which momentum equations for individual Fourier modes can be analyzed term by term. Named Fourier-Averaged Navier-Stokes (FANS) equations, the method derives individual momentum equations at each timescale in terms of Fourier modes for pressure and velocity. The method aims to analyze the contribution of pressure, diffusive, convective and unsteady momentum fluxes to budgets at specific timescales. This strategy is analogous to analysing the turbulent kinetic energy transport terms that arise when studying the Reynolds-Averaged Navier Stokes (RANS) equations. The current research proposes that analysing these momentum budgets is useful to directly identifying the processes that govern the flow physics. By evaluating the convective coupling between Fourier modes, this method also allows for direct comparison of triadic interactions to other momentum fluxes. The method relies on well-known data processing techniques (Fourier decomposition and numerical gradient approximation) which aids the physical interpretation of the results. To evaluate the application of FANS for analysing the momentum balance of complex flows, three case studies are considered: that of periodic flows in (i) the wake a square cylinder and (ii) swirling jet; as well as the cyclical but nonperiodic flow around two cylinders arranged side-by-side. Here, periodic flows refer to those that repeat regularly in time, while cyclic flows refer to any case with recurring patterns, whether they are regular or irregular. The square cylinder serves to illustrate the methodology on a simple case. The jet flow shows the application of the method in a case where there are a large number of energetically important frequencies and additional complexity due to the swirl. Finally, the dual cylinder case illustrates the application when the frequency signature does not only consist of a single dominant frequency and harmonics. A comparison to the BMD analysis is also provided.

The structure of the paper is as follows: first the derivation of the FANS equations will be detailed and the terms will be labeled and interpreted. A brief overview of the Bispectral Mode Decomposition will also be presented for context. Subsequently, case studies will be conducted and discussed. Finally, a few concluding remarks will be made.

\section{Methods}
Overviews of the methods used in this paper are presented in two parts. We begin by deriving the FANS equations and providing the physical interpretation of different terms therein. Then, we proceed with a brief review of the Bispectral Mode Decomposition and its derivation, which is referred to later on in the case studies.

\subsection{Fourier-Averaged Navier-Stokes equations}
FANS assumes a Fourier series representation to the solution of the governing equations to arrive at a decomposition of the momentum equations. The goal of this decomposition is to elucidate the momentum transfer processes at different timescales. Starting from the incompressible, nondimensionalized Navier-Stokes equations:
\begin{equation}
    \frac{\partial \mathbf{u}}{\partial t^*} + \mathbf{u} \cdot \nabla \mathbf{u} = -\nabla p + \frac{1}{Re} \nabla^2 \mathbf{u},
\end{equation}
where $\mathbf{u}$ is the velocity field and (represented in bold as a vector quantity) and $p$ is the pressure field, we assume these fields satisfy Fourier series:
\begin{equation*}
    \mathbf{u} = \sum_{m \in \mathcal{Z}} \bhat{u}^m e^{j\omega m t}
\end{equation*}
and
\begin{equation*}
    p = \sum_{m \in \mathcal{Z}} \hat{p}^m e^{j\omega m t}.
\end{equation*}
Here, $j=\sqrt{-1}$ and the spatially varying coefficients of the Fourier series ($\bhat{u}^m, \hat{p}^m$) are called the velocity and pressure modes. Inserting this ansatz into the Navier-Stokes equations returns
\begin{equation*}
    \frac{\partial}{\partial t} \sum_{m} \bhat{u}^m e^{j\omega m t} + \sum_{m} \left(\bhat{u}^m e^{j\omega m t}\right) \cdot \nabla \sum_{n} \bhat{u}^n e^{j\omega n t} = -\nabla \sum_{m} \hat{p}^m e^{j\omega m t} + \frac{1}{Re} \nabla^2 \sum_{m} \bhat{u}^m e^{j\omega m t}.
\end{equation*}
As differentiation is linear, the derivative operators may be brought into the summations:
\begin{equation*}
    \sum_m \bhat{u}^m \frac{\partial}{\partial t} e^{j\omega m t} + \sum_m \sum_n (\bhat{u}^m \cdot \nabla \bhat{u}^n) e^{j\omega m t} e^{j\omega n t} = \sum_m \left(-\frac{1}{Re} \hat{p}^m + \frac{1}{Re}\nabla^2 \bhat{u}^m\right)e^{j\omega m t}.
\end{equation*}
By exploiting the orthogonality of the expansion coefficients, $\exp\left(j\omega mt\right)$, an equation for a momentum balance for a particular mode $(\bhat{u}_k)$ can be obtained. Specifically, the balance at the timescale corresponding to a particular mode can be expressed in terms of that mode and the other Fourier modes:
\begin{equation*}
    j2 \pi f k \bhat{u}^k
     +  \mathbf{U}\cdot\nabla \bhat{u}^k + \bhat{u}^k \cdot\nabla \mathbf{U}
     = 
    -\nabla \hat{p}^k 
    + \frac{1}{Re} \nabla^2 \bhat{u}^k
    - \sum_{n\neq 0, k} \bhat{u}^n \cdot \nabla \bhat{u}^{k-n}.
\end{equation*}

\begin{table}
    \centering
    \begin{tabular}{c|c}
        Term & Interpretation \\
        \hline
         $j\omega k \bhat{u}^k$ & Unsteady \\
         $\mathbf{U}\cdot\nabla \bhat{u}^k + \bhat{u}^k \cdot\nabla \mathbf{U}$ & Mean flow convection \\
         $-\nabla \hat{p}^k$ & Pressure \\
         $\frac{1}{Re} \nabla^2 \bhat{u}^k$ & Viscosity/Diffusion \\
         $\sum_{n\neq 0, k} \bhat{u}^n \cdot \nabla \bhat{u}^{k-n}$ & Inter-mode convection
    \end{tabular}
    \caption{Interpretation of individual FANS terms.}
    \label{tab:interpretation}
\end{table}
Here, $\mathbf{U}$ represents the mean flow. The physical interpretation of the terms are detailed in table \ref{tab:interpretation}. Convective interactions between phenomena at different frequencies is of particular interest as they are triadic in nature. For convenience for the rest of this paper, the inter-mode convection term will be written as:
\begin{equation}\label{eq:chi}
    \chi[\bhat{u}^k] = \sum_{n\neq 0, k} \bhat{u}^n \cdot \nabla \bhat{u}^{k-n}.
\end{equation}
We refer to these terms as the Fourier stresses by way of analogy to the Reynolds stresses. It suggests a series of kinetic energy equations for each timescale, similar to the turbulent kinetic energy equation that arises due to the closure problem in the Reynolds-averaged Navier-Stokes equations. 

The momentum equation can be written to highlight the processes that affect the rate of change of a mode:
\begin{equation}\label{eq:FANS}
    j2\pi kf \bhat{u}^k
    =- \mathbf{U}\cdot\nabla \bhat{u}^k - \bhat{u}^k \cdot\nabla \mathbf{U} 
    -\nabla \hat{p}^k 
    + \frac{1}{Re} \nabla^2 \bhat{u}^k
    - \chi[\bhat{u}^k].
\end{equation}
The left hand side represents the contribution to the unsteady fluctuations at a point in space at a timescale $(kf)^{-1}$. Note that the unsteady term (UT) at frequency $f_k=kf$ is a scalar multiple of the mode $\bhat{u}^k$.

In practice, this process will be applied on experimental or simulation data that are discretely sampled. As a result, the modes are constructed from a discrete Fourier transform of the $N$ snapshots, $\mathbf{u}_n$:
\begin{equation}
    \bhat{u}^k = \sum_{n=0}^N \mathbf{u}_n e^{-j2\pi kf n},
\end{equation}
where $f=1/N$. The expansion coefficients ($e^{-j2\pi kfn}$) have a useful orthogonality property that allows us to find the momentum transport processes for each mode. Using a discrete approximation of the momentum equations, we can find the following discrete version of equation (\ref{eq:FANS}):
\begin{equation}\label{FANS-discrete}
    j2\pi kfn \bhat{u}^k = - C[\mathbf{U}, \bhat{u}^k] - C[\bhat{u}^k, \mathbf{U}] - \mathbf{G}[\hat{p}^k] + \frac{1}{Re} L[\bhat{u}^k] - \chi[\bhat{u}^k],
\end{equation}
where $C[\cdot, \cdot]$ is a discrete convection operator, $\mathbf{G}[\cdot]$ is a discrete gradient, $L[\cdot]$ is a discrete Laplacian operator, and $\chi[\bhat{u}^k] = \sum_{n\neq 0, k} C[\bhat{u}^n, \bhat{u}^{k-n}]$. In the case of simulation data, it is recommended that these operators are selected to be the same as the ones used to set up the system of equations. Using the same operators is advantageous as it does not introduce new discretization errors.

Interpretation of the terms in the FANS formulation reveals how flow dynamics can be explored. In (\ref{FANS-discrete}), the unsteady term - and thus the mode - can be isolated and calculated as a balance of forces due to convection, diffusion, and pressure. This separates the contributions of each force to the momentum balance and allows comparison of their magnitudes. Likewise, analysing phases of the terms shows whether the forces are enhancing or resisting the unsteady fluctuations (UT) locally. Forces that are in-phase increase the magnitude of the UT, while forces that are out of phase resist it. Both location and frequency data can be used to associate forces with the physics. For instance, the locations and timescales associated with a particular structure can be used to identify which forces are significant in the evolution of that structure. If there are several frequencies associated with a particular structure, a summation of those frequencies can be used to investigate the relationships between forces that act on it.

FANS may also be used to identify nonlinear interactions between modes directly by estimating the convective term that drives them. Namely, the term $\bhat{u}^n\cdot\nabla \bhat{u}^{k-n}$ represents the driving force of a triadic interaction between modes $k$, $n$, and $k-n$. The consequences of this representation will be investigated in later sections.

Two methods can be used to calculate the terms from flow field data. The first option is to calculate the Navier-Stokes terms on the original snapshots and then apply the Discrete Fourier Transform (DFT) on each set of terms. The other option is to apply the DFT to the snapshots and then calculate the FANS terms from the modes. For the UT, diffusion, pressure, and mean-flow convection terms, these two methods are equivalent. However, the first method allows the Fourier stresses to be calculated quickly by calculating the derivatives in the time domain, whereas the second method requires calculating a sum of $N$ convection terms in the frequency domain for each of the $N$ modes.

The derivation of FANS that results in (\ref{FANS-discrete}) implies a single, long-time box window including all available data. As a result, the standard rules apply with regards to sampling frequency and window size. The analysis window must be sufficiently large to allow for good frequency resolution.
Furthermore, the use of DFT in this case indicates that FANS is highly useful in cases where energy is concentrated in limited ranges of frequencies.

\subsection{Bispectral Mode Decomposition}
Bispectral Mode Decomposition, introduced by \citet{schmidt-bmd}, is a modal decomposition used to identify the locations of triadic interactions. The method specifically aims to extract modes that ``exhibit quadratic phase coupling over extended portions of the flow field". This is done by means of a spatial integration. BMD is a data-based method that uses the phase coupling relationship as a proxy to identify interactions between modes, which contrasts to FANS, which uses physical arguments to identify interactions.

Interested readers are referred to the original derivation in \citet{schmidt-bmd}, the basic overview and results of which are included here for completeness.  The method is derived to be general for any process that can be described as a set of quadratically-nonlinear partial differential equations. The quantity of interest for BMD is a vector of complex expansion coefficients $\mathbf{a}_{k,l}=a_{k,l}^j$ for each triad ($k$, $l$, $k+l$) that maximizes the quantity
\begin{equation}
    \mathbf{a}_1 = \argmax_{\|\mathbf{a}_{k,l}\| = 1} \big\rvert \mathbf{a}_{k,l}^H \bhat{Q}^H_{k\circ l} \mathbf{W} \bhat{Q}_{k + l} \mathbf{a}_{k,l} \big\rvert,
\end{equation}
such that
\begin{equation}
    \lambda_1 = \mathbf{a}_1^H \bhat{Q}^H_{k\circ l} \mathbf{W} \bhat{Q}_{k + l} \mathbf{a}_1.
\end{equation}
The superscript $\square^H$ represents the conjugate transpose of a vector or matrix. The quantity $\lambda_1$ can be interpreted as representing a maximized bicorrelation for a particular triad. $\mathbf{W}$ is a weight matrix that is used to approximate spatial integration. $\bhat{Q}_{k\circ l}=[\bhat{q}_k^0\circ\bhat{q}_l^0 ... \bhat{q}_k^n\circ\bhat{q}_l^n]$ (where $\bhat{q}_k \circ \bhat{q}_l$ is the elementwise product of two vectors) and $\bhat{Q}_{k + l}=[\bhat{q}_{k+l}^0 ... \bhat{q}_{k+l}^n]$ are matrices formed from the Fourier modes ($\bhat{q}^i_k$) of sequences of snapshots that have been split into $n$ blocks.

Once the maximizing expansion coefficients have been calculated, the desired modes of BMD can be obtained:
\begin{equation*}
    \phi_{k+l} = \mathbf{Q}_{k+l}\mathbf{a}_1,
\end{equation*}
\begin{equation*}
    \phi_{k\circ l} = \mathbf{Q}_{k\circ l}\mathbf{a}_1,
\end{equation*}
\begin{equation*}
    \psi_{k,l} = \rvert \phi_{k+l} \circ \phi_{k\circ l} \rvert.
\end{equation*}
$\phi_{k+l}$ is named the ``bispectral mode", and represents the resultant mode of the triadic interaction. $\phi_{k\circ l}$ is the ``cross-frequency" mode and represents the effect of the input modes. $\psi_{k,l}$ is deemed the ``interaction map" and shows the magnitude of the local bicorrelation between the triads. The values of $\lambda_1$ for all triads $[k, l, k+l]$ is known as the mode bispectrum and represents the relative magnitudes of interactions, evaluated globally. 

\section{Case studies}
Three case studies are considered to describe the application of FANS and the physical interpretation of its results. Two periodic (regularly repeating) cases, flow over a square cylinder and a swirling jet impinging on a wall, and a cyclic but nonperiodic (irregularly repeating) case of two side by side cylinders are analysed for this purpose. The data is obtained from direct numerical simulations completed in OpenFOAM and the results are validated against existing results in the literature. The case studies are presented in order of increasing complexity, starting with the simple, 2D case of flow around a square cylinder where the fluctuations are well described by the mean and two most energetic Fourier modes, consisting of the fundamental frequency and second harmonic. Then, an axisymmetric swirling jet is considered, as there is an increase in the level of complexity due to the increase in the number of relevant frequencies and additional velocity component. Finally, irregular vortex shedding behind two adjacent cylinders is considered to evaluate the application of FANS for characterizing flows with broadband velocity and pressure signals. This case retains cyclic characteristics resulting in broadband spectral accumulations about particular frequencies.

\subsection{Square cylinder}
\begin{figure}
    \centering
    \includegraphics[scale=0.7]{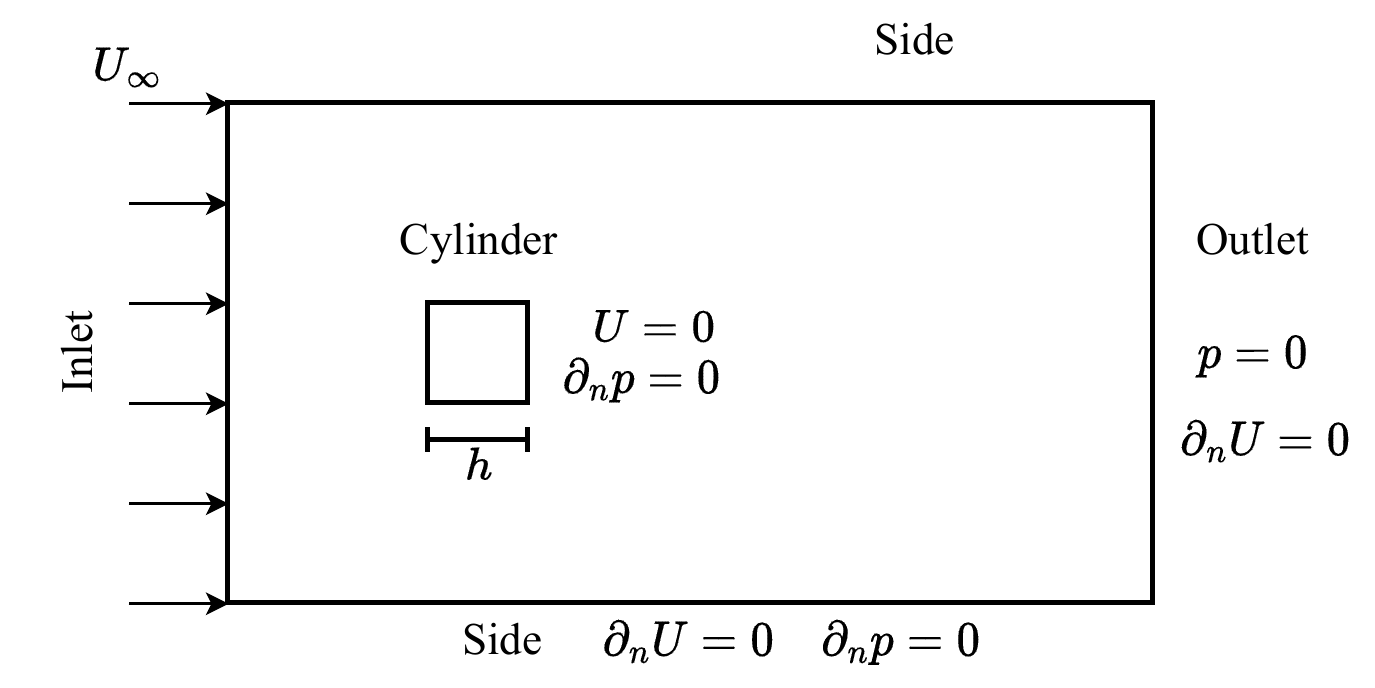}
    \caption{2D Simulation domain of a square cylinder (not to scale).}
    \label{fig:square-domain}
\end{figure}
We begin with the simple case of flow over a 2D square cylinder at Reynolds number of 100. The flow is periodic, exhibiting a dominant frequency and harmonics. This makes this a natural starting point for illustrating FANS-based analysis. The 2D simulation mesh is based on parameters of \citet{bai-cylinder}, against which the present simulations have been validated. A visualization of the domain and boundary conditions is shown in figure \ref{fig:square-domain}. 

\begin{figure}
    \centering
    \includegraphics[scale=0.7]{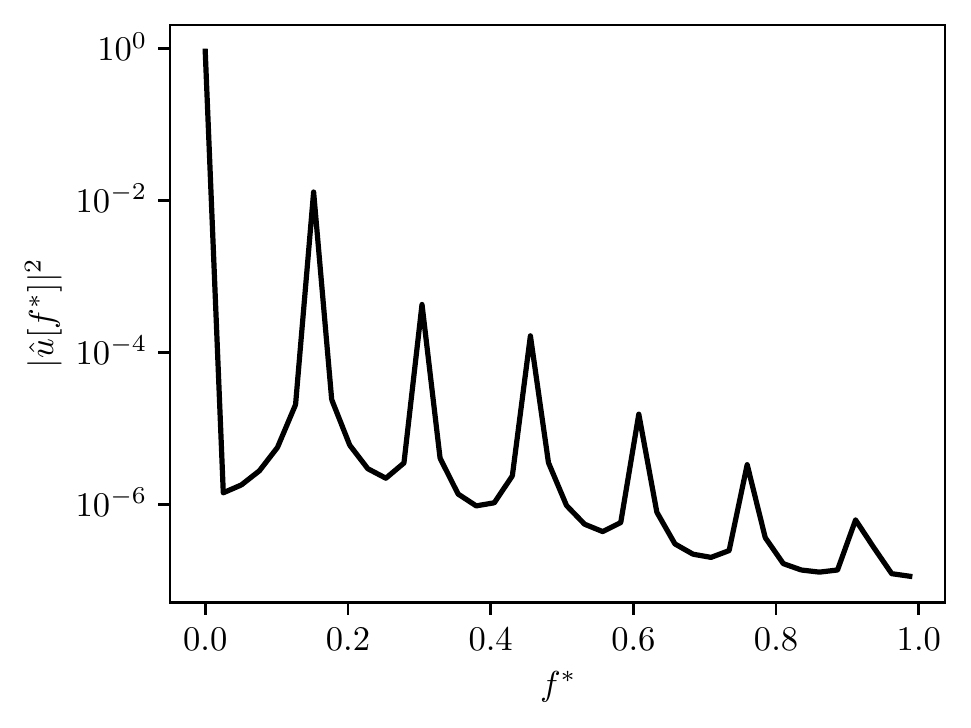}
    \caption{Velocity spectrum in the wake of a square cylinder at a $Re=100$. The fundamental frequency is $f^*=0.15$.}
    \label{fig:square-spectrum}
\end{figure}
The wake behind the cylinder is characterized as a classic periodic (Karman) vortex shedding process, described by \citet{williamson}. These vortices grow due to diffusion and diverge from the centreline as they are convected downstream. This shedding results in a velocity spectrum characterized by distinct peaks that diminish rapidly in magnitude as the mode number increases, as seen in figure \ref{fig:square-spectrum}. As a result, only the mean and modes at the fundamental frequency and second harmonic are considered. These modes have a maximum magnitude of 1.3$U_\infty$, 0.3$U_\infty$, and 0.07$U_\infty$, respectively. Here, FANS is used to explore the momentum transportation associated with the vortex street as well as the interactions between newly-formed vortices immediately behind the cylinder.

\begin{figure}
    \centering
	\begin{minipage}{0.5\textwidth}
		\centering
		\subcaptionbox{\hspace*{-1.75em}}{%
			\hspace{-0.25in}	\includegraphics[scale=0.7]{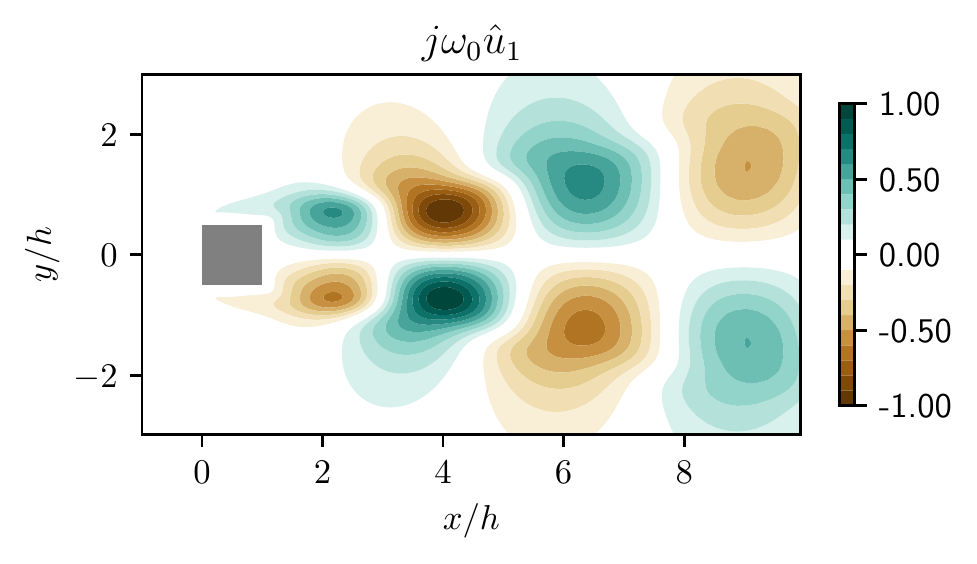}%
		}\qquad
	\end{minipage}\hfill
	\begin{minipage}{0.5\textwidth}
		\centering
		\subcaptionbox{\hspace*{-1.75em}}{%
			\hspace{-0.25in}	\includegraphics[scale=0.7]{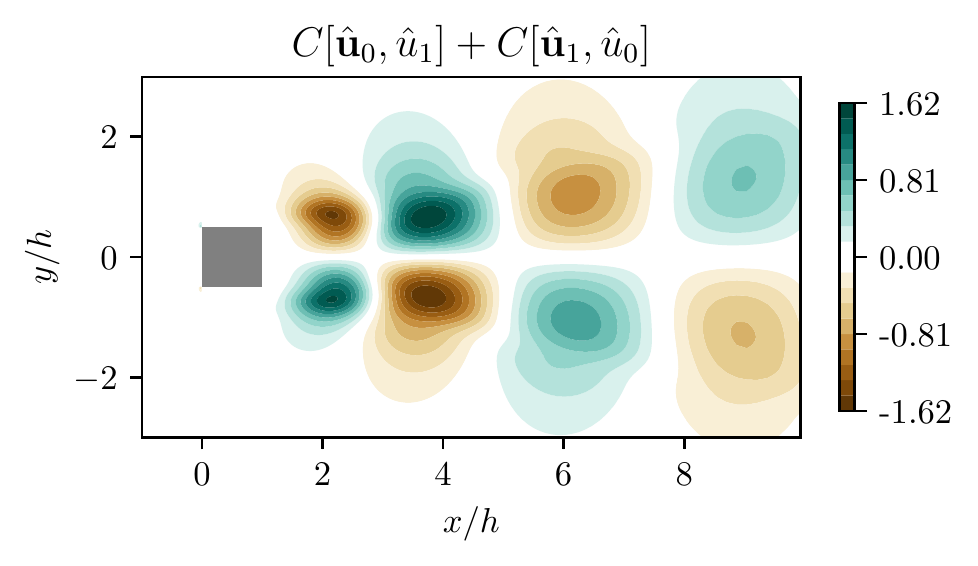}%
		}\qquad
	\end{minipage}\\
	\begin{minipage}{0.5\textwidth}
		\centering
		\subcaptionbox{\hspace*{-1.75em}}{%
			\hspace{-0.25in}	\includegraphics[scale=0.7]{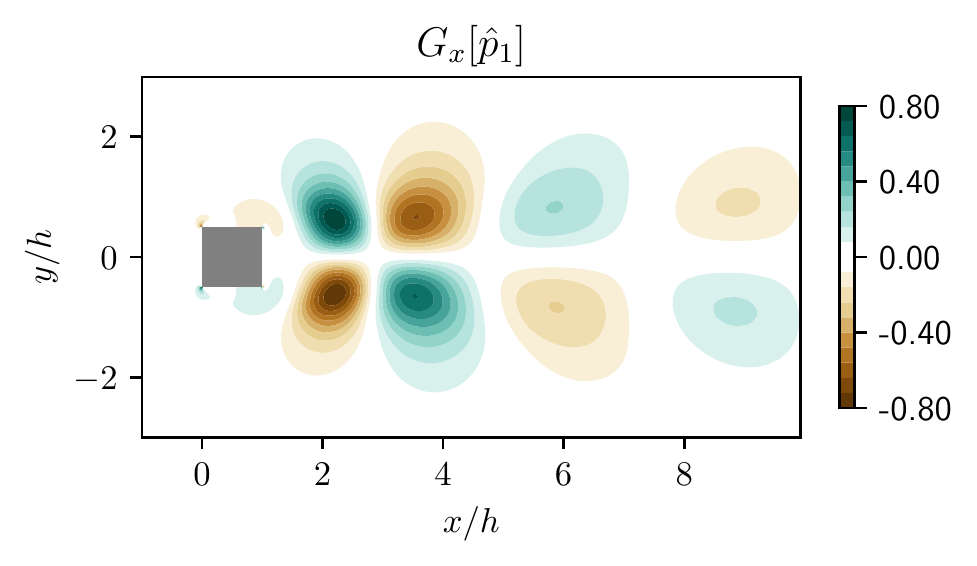}%
		}\qquad
	\end{minipage}\hfill
	\begin{minipage}{0.5\textwidth}
		\centering
		\subcaptionbox{\hspace*{-1.75em}}{%
			\hspace{-0.25in}	\includegraphics[scale=0.7]{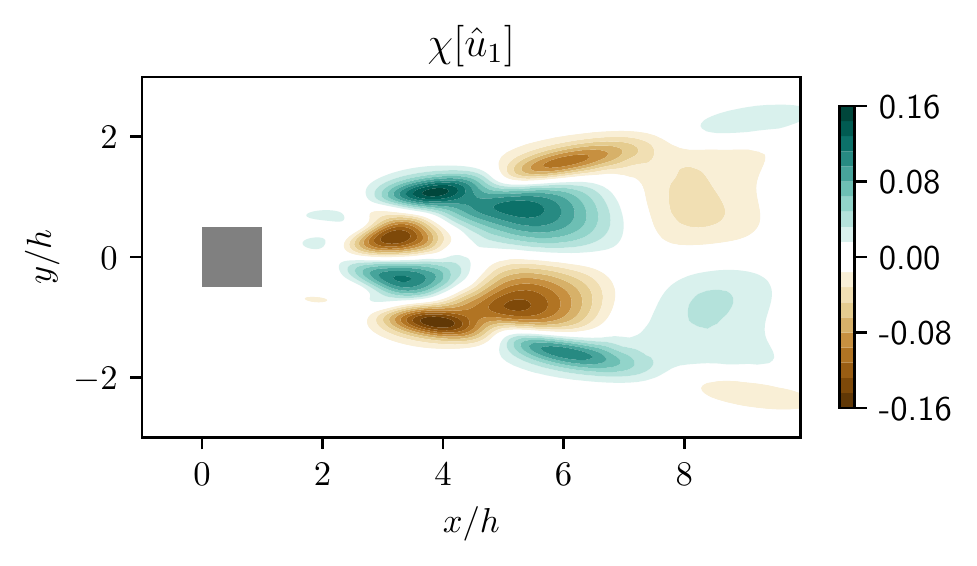}%
		}\qquad
	\end{minipage}
    \caption{Streamwise momentum terms for mode 1 (at the fundamental frequency) in the wake of a square cylinder at $Re=100$: (a) UT (b) mean-flow convection (c) pressure gradient and (d) Fourier stress terms.}
    \label{fig:fans-square-CT1}
\end{figure}

Figure \ref{fig:fans-square-CT1} shows the real values of FANS terms at the fundamental frequency ($f^*=fh/U_\infty=0.15$) for the streamwise velocity, $\hat{u}^1$. The data are represented as a fraction of the maximum of the UT in the domain at the same frequency and direction, $T^1_{max}$:
\begin{equation}
    T^k_{max} = \max_{\vec{x}}(\rvert j2\pi fk \hat{u}^k(\vec{x})\rvert).
\end{equation}
For example, the value displayed in figure \ref{fig:fans-square-CT1}(a) is $j2\pi f^*k \hat{u}^1/T^1_{max}$.  By applying this same normalization to each flux at a given frequency, the relative contribution of each term can be compared. 
Figure \ref{fig:fans-square-CT1}(a) shows UT ($j2\pi f^*\hat{u}^1$) of the fundamental frequency in the streamwise direction. Contours of the UT are associated with the large-scale fluctuations of the vortices. Figure \ref{fig:fans-square-CT1}(b) shows the mean flow convection term $C[\mathbf{U}, \hat{u}^1] + C[\mathbf{\hat{u}}^1, {U}]$. Figure \ref{fig:fans-square-CT1}(c) shows the pressure term  and figure \ref{fig:fans-square-CT1}(d) shows the Fourier stresses. The viscous dissipation is of negligible magnitude and it is not shown here for brevity. Downstream of the cylinder (past $x=8h$), the only force with significant magnitude is the mean-flow convection. This represents the region where fully-formed vortices are convected downstream. In FANS terms, the mean-flow convection is nearly balanced by the UT. 

Immediately behind the cylinder, there are pressure gradients of large magnitude, which are related to the vortex formation. This is contrary to the downstream wake region. This way of comparing the magnitude of momentum terms identifies the critical forces by region. The pressure gradient in this region is in-phase with the UT and out-of-phase with the mean-flow convection. From this phase relationship, the pressure gradient reduces the overall magnitude of the velocity fluctuations caused by movement of the vortices.

\begin{figure}
    \centering
	\begin{minipage}{0.5\textwidth}
		\centering
		\subcaptionbox{\hspace*{-1.75em}}{%
			\hspace{-0.25in}	\includegraphics[scale=0.7]{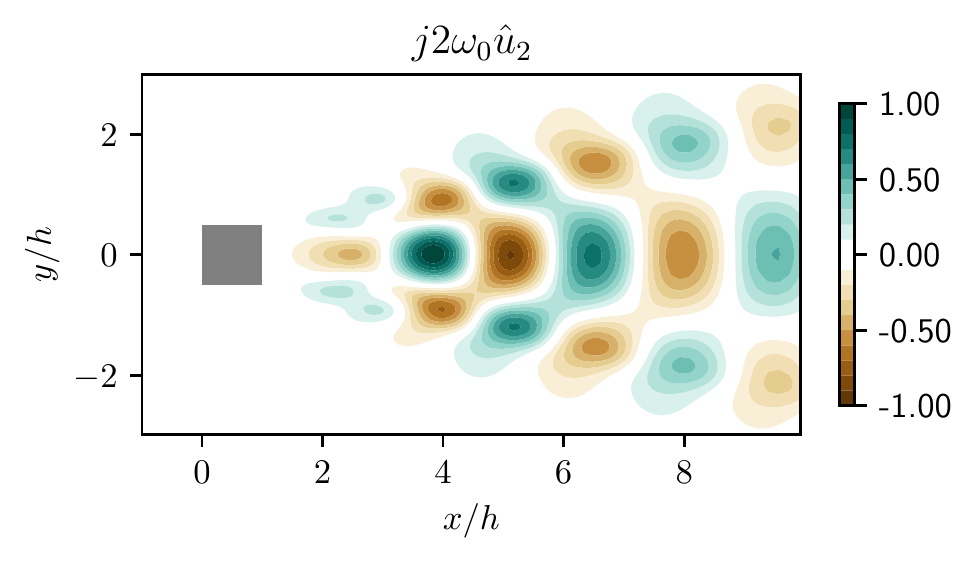}%
		}\qquad
	\end{minipage}\hfill
	\begin{minipage}{0.5\textwidth}
		\centering
		\subcaptionbox{\hspace*{-1.75em}}{%
			\hspace{-0.25in}	\includegraphics[scale=0.7]{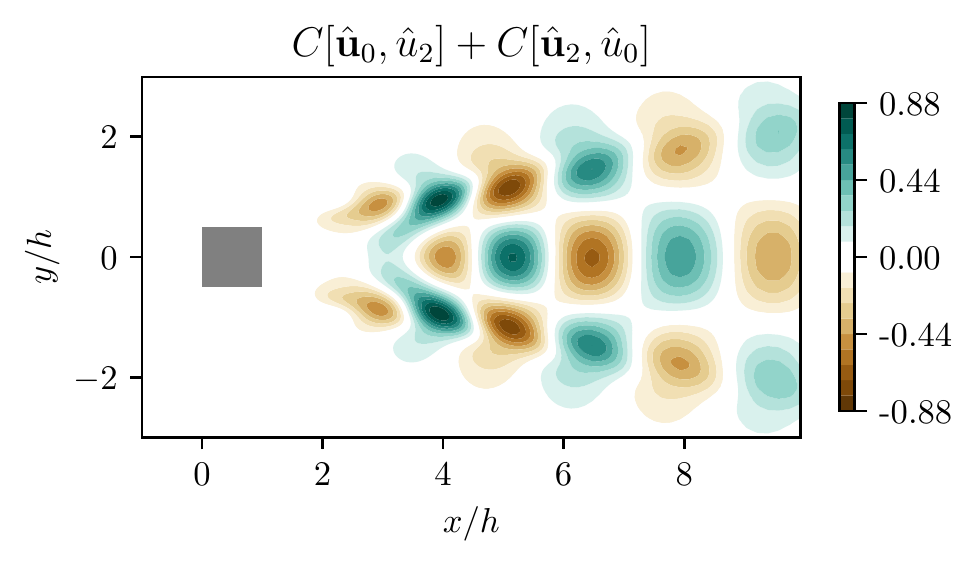}%
		}\qquad
	\end{minipage}\\
	\begin{minipage}{0.5\textwidth}
		\centering
		\subcaptionbox{\hspace*{-1.75em}}{%
			\hspace{-0.25in}	\includegraphics[scale=0.7]{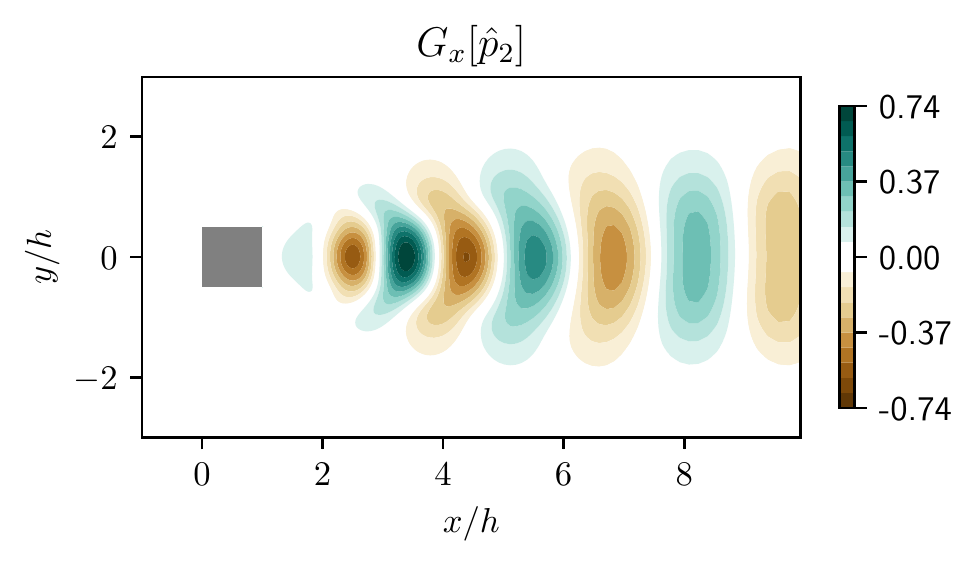}%
		}\qquad
	\end{minipage}\hfill
	\begin{minipage}{0.5\textwidth}
		\centering
		\subcaptionbox{\hspace*{-1.75em}}{%
			\hspace{-0.25in}	\includegraphics[scale=0.7]{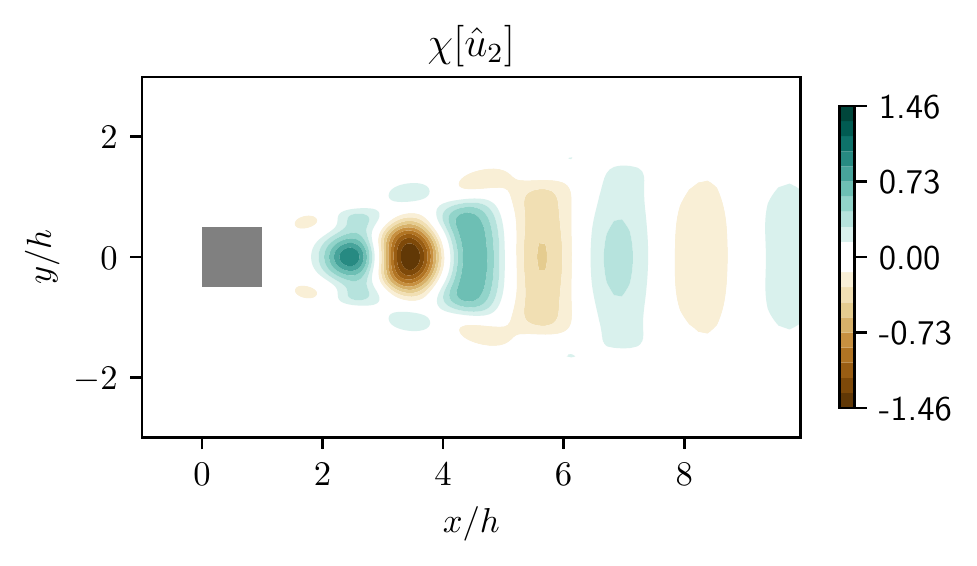}%
		}\qquad
	\end{minipage}
    \caption{Real part of streamwise (a) unsteady (b) mean-flow convection (c) pressure gradient and (d) Fourier stress terms for the second harmonic of vortex shedding over a square cylinder at $Re=100$.}
    \label{fig:fans-square-CT2}
\end{figure}

The low magnitude of the Fourier stress term indicates that the inter-frequency interactions are not as important as the other fluxes to the momentum balance at the fundamental frequency. This is consistent with the findings of other methods, such as linear stability or the Self Consistent Method \citep{meliga_2016}. These stresses are generated by interaction of modes $\hat{u}_1$ and $\hat{u}_2$ in the form $\bhat{u}_{-1}\cdot \nabla \hat{u}_2$. The significantly reduced magnitude of the harmonic (mode 2) with respect to mode 1 results in the low magnitude of these stresses. A representation of mode 2 is shown in figure \ref{fig:fans-square-CT2}(a) in the form of the UT. The magnitude of the harmonic is elevated at the centreline and in two branches diverging from the centreline, which correspond to the edges of each vortex track.

The other momentum components in the streamwise direction for the second harmonic are also shown in figure \ref{fig:fans-square-CT2}(b), (c), and (d). The relationships between different momentum fluxes are the same as they are in the fundamental frequency. Away from the axis and farther downstream in the wake, the unsteady and convection terms are nearly balanced. Behind the cylinder, the pressure term is significant and in-phase with the velocity fluctuations.

The Fourier stresses play a more critical role in affecting the balance at the harmonic in comparison to the fundamental frequency. The convective interactions represented by this term are significant to the momentum transport at this frequency. This momentum flux is strong and symmetric about the wake centreline, and dissipates rapidly moving downstream. To better understand this momentum flux, the constituent terms of the Fourier stresses are analysed. These terms arise from the expansion of $\chi[\hat{u}^2]$ from (\ref{eq:chi}): 
\begin{equation}
    \chi[\hat{u}^2] = ... + \hat{u}^{-1}\partial_x \hat{u}^3 + \hat{v}^{-1}\partial_y \hat{u}^3 + \hat{u}^{1}\partial_x \hat{u}^1 + \hat{v}^{1}\partial_y \hat{u}^1 + \hat{u}^{3}\partial_x \hat{u}^{-1} + \hat{v}^{3}\partial_y \hat{u}^{-1} + ...
\end{equation}
For this flow, terms other than $\hat{u}^{1}\partial_x \hat{u}^1$ and $\hat{v}^{1}\partial_y \hat{u}^1$ are negligible. This is similar to the results of \citet{cylinder-scm} for a cylinder wake. Due to their influence on the momentum flux of velocity mode 2, these terms can be interpreted as the driving force of triadic interactions between the fundamental frequency, itself, and the second harmonic.
\begin{figure}
    \centering
	\begin{minipage}{0.5\textwidth}
		\centering
		\subcaptionbox{\hspace*{-1.75em}}{%
			\hspace{-0.25in}	\includegraphics[scale=0.7]{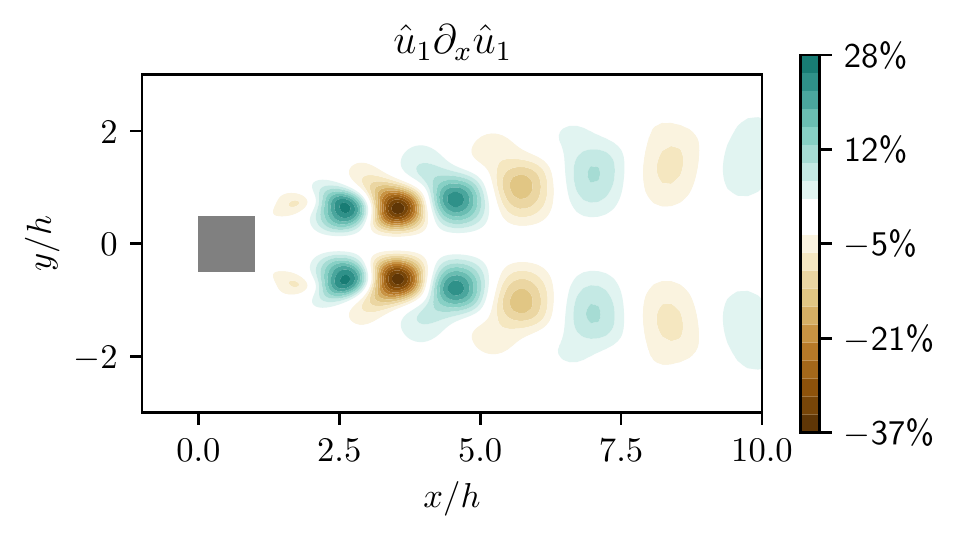}%
		}\qquad
	\end{minipage}\hfill
	\begin{minipage}{0.5\textwidth}
		\centering
		\subcaptionbox{\hspace*{-1.75em}}{%
			\hspace{-0.25in}	\includegraphics[scale=0.7]{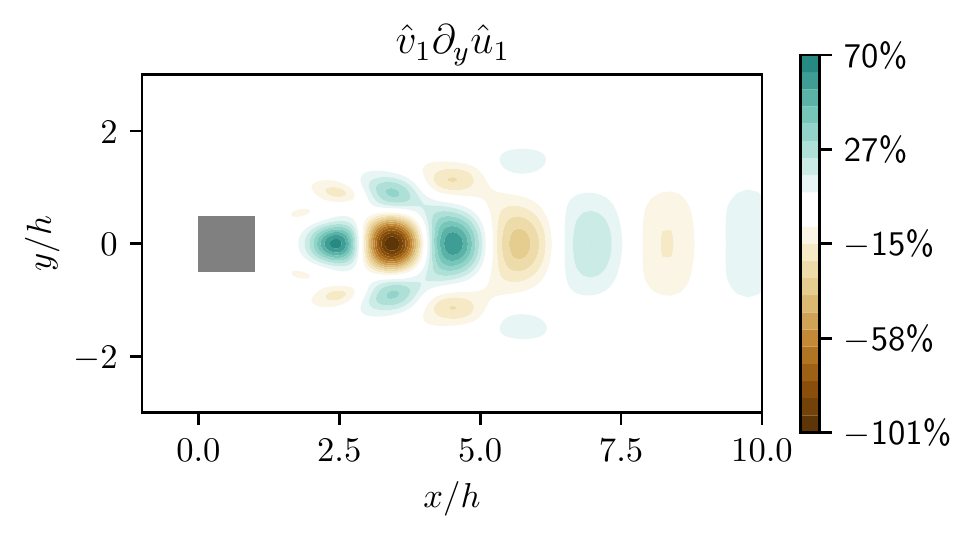}%
		}\qquad
	\end{minipage}
    \caption{Constituent terms of the Fourier stresses ($\chi[\hat{u}^2]$) in the streamwise direction at the second harmonic for a square cylinder at $Re=100$. (a) first term (b) second term. Only the real parts of the stresses are shown.}
    \label{fig:fstress-components}
\end{figure}
The real parts of these terms are depicted in figure \ref{fig:fstress-components}. The values are normalized to the maximum Fourier stress in the domain ($\chi[\hat{u}^2]$). The first term ($\hat{u}^{1}\partial_x \hat{u}^1$) roughly follows the track of the vortex street. This flux is strongest near the vortex formation region, and then dissipates as the vortices separate and lose strength downstream. Hence, this term is a result of convective momentum transport between streamwise velocity fluctuations within an individual vortex. Peaking at roughly 40\% of the magnitude of $\chi[\hat{u}^2]$, this component makes a significant contribution to the momentum flux, especially in the vortex formation region. However, the other term ($\hat{v}^{1}\partial_y \hat{u}^1$) in figure \ref{fig:fstress-components}(b) has a more significant effect. This term attains 100\% of the maximum Fourier stress in the vortex formation region. Thus, this is the primary term of the Fourier stresses at the second harmonic of the streamwise velocity. The presence of this momentum transport at the vortex formation region indicates that the second harmonic is a product of interactions between pairs of counter-rotating vortices formed off of the top and bottom surfaces of the cylinder. Namely, these vortices are in close proximity, and the resulting contact modifies their shape. Interactions between these vortices will occur every half-cycle, hence the appearance of this process in the momentum equation at the second harmonic. This shows how interactions at the fundamental frequency result in increased energy content of the second mode. 

\begin{figure}
    \centering
    \includegraphics[scale=0.8]{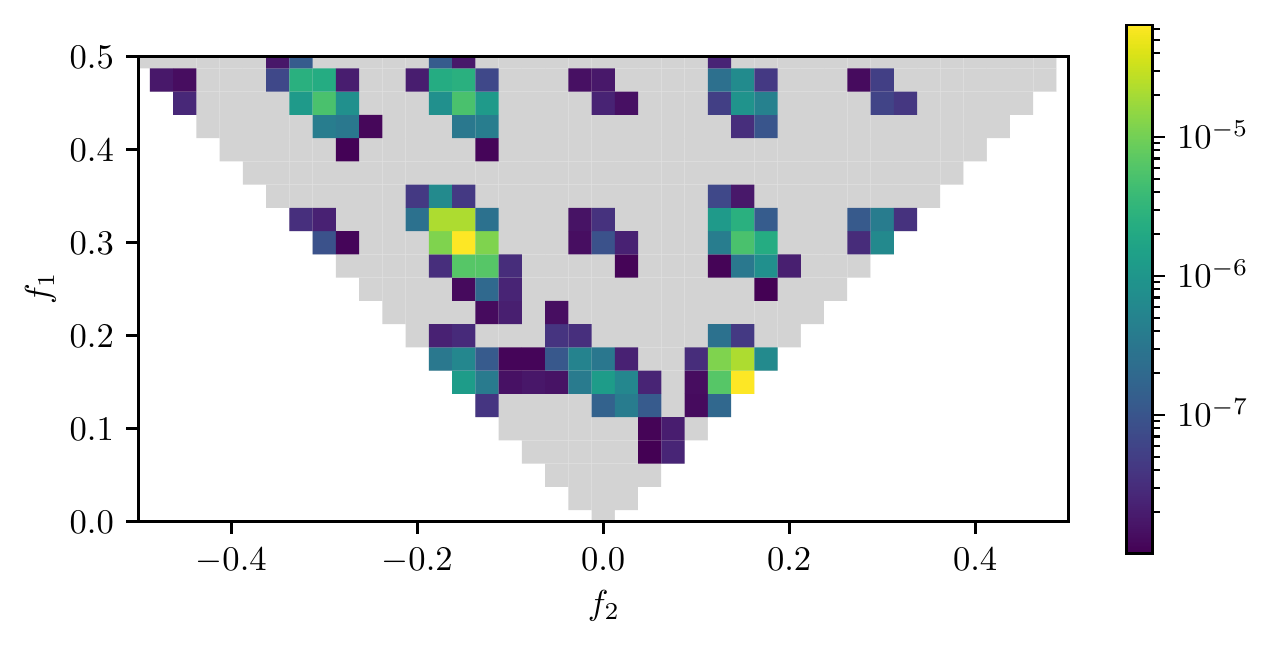}
    \caption{BMD mode bispectrum of square cylinder at $Re=100$.}
    \label{fig:bmd-spec-square}
\end{figure}

FANS analysis can directly deduce interactions between timescales, as discussed above. These interactions can also be deduced using BMD, however this method can be more cumbersome. For this flow, the BMD modes and bispectrum have been calculated for two regions of the BMD search space described by \citet{schmidt-bmd}. Two regions are considered as there are symmetries in the results due to the periodicity and real-valued flow field data fed into the Fourier decomposition. Results for these regions, consisting of sums and differences of two base frequencies ($f_1$ and $f_2$) are presented in figure \ref{fig:bmd-spec-square}. The BMD mode bispectrum shows the value of $\lambda_1$ for each triad, representing the maximized bicorrelation between each triplet of modes. This spectrum shows the cascade starting from mode 1 at the shedding frequency of $f^*=0.15$, which continues through higher modes at the harmonics of this frequency. This is seen in the local maxima of the bispectrum at $(f^*_1=0.15, f^*_2=0.15)$ which is correlated to $f^*_3=0.3$. This interaction, and ensuing interactions between the fundamental frequency and the harmonics, results in the series of peaks in the mode bispectrum that attenuate with increasing frequencies $f^*_1$ and $f^*_2$. This is similar to the results of \citet{schmidt-bmd} for the case of a circular cylinder. This cascade is represented in FANS by the Fourier stress terms in figures \ref{fig:fans-square-CT1}(d) and \ref{fig:fans-square-CT2}(d).

\begin{figure}
    \centering
    \includegraphics[scale=0.11]{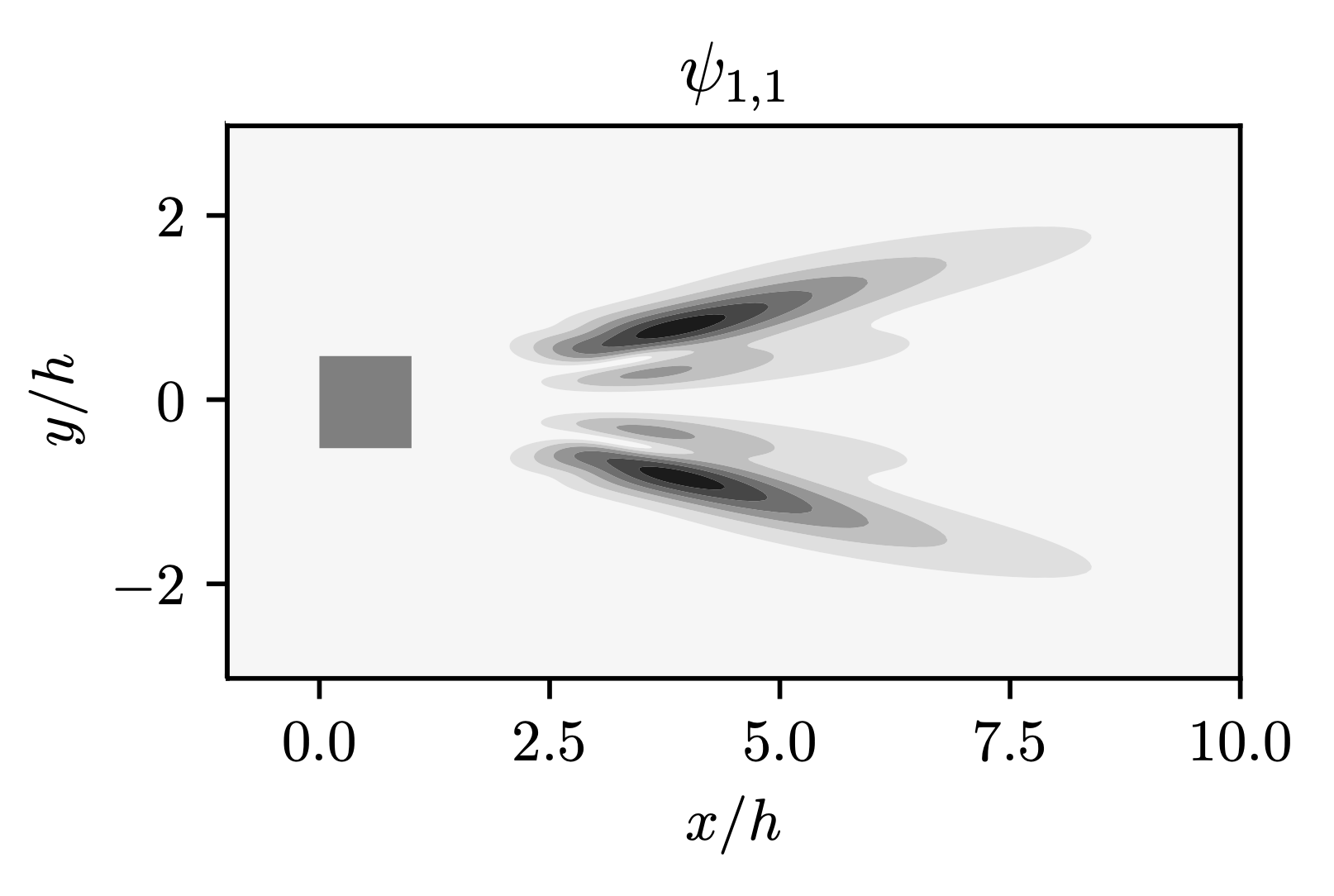}
    \caption{BMD interaction map of triad $\{f_1, f_1, f_2\}$ in the streamwise direction for a square cylinder at $Re=100$.}
    \label{fig:bmd-square-u112}
\end{figure}

Components of the bispectral mode ($\phi_{1+1}$) and interaction map ($\psi_{1,1}$) that correspond to the triad $(f_1=0.15, f_2=0.15, f_3=0.30)$ are shown in figure \ref{fig:bmd-square-u112} for the streamwise direction. The interaction maps clearly show the interactions between the fundamental frequency and second harmonic that occur in the Karman wake behind the cylinder. This interaction of the triad is similar to the streamwise Fourier stresses on the second harmonic, shown in figure \ref{fig:fstress-components}(a). 
For the square cylinder case, the local interactions detected by BMD are also captured by FANS. An added benefit of FANS analysis is that those interactions can be related to other momentum transport terms and the time dependence of the flow, as seen in figures \ref{fig:fans-square-CT1} and \ref{fig:fans-square-CT2}. This case study supports that FANS is a simpler method that produces insights into flow features and their interactions for for periodic wakes characterised by a single fundamental frequency and its harmonics.

\subsection{Swirling Jet}
\begin{figure}
    \centering
    \includegraphics[scale=0.75]{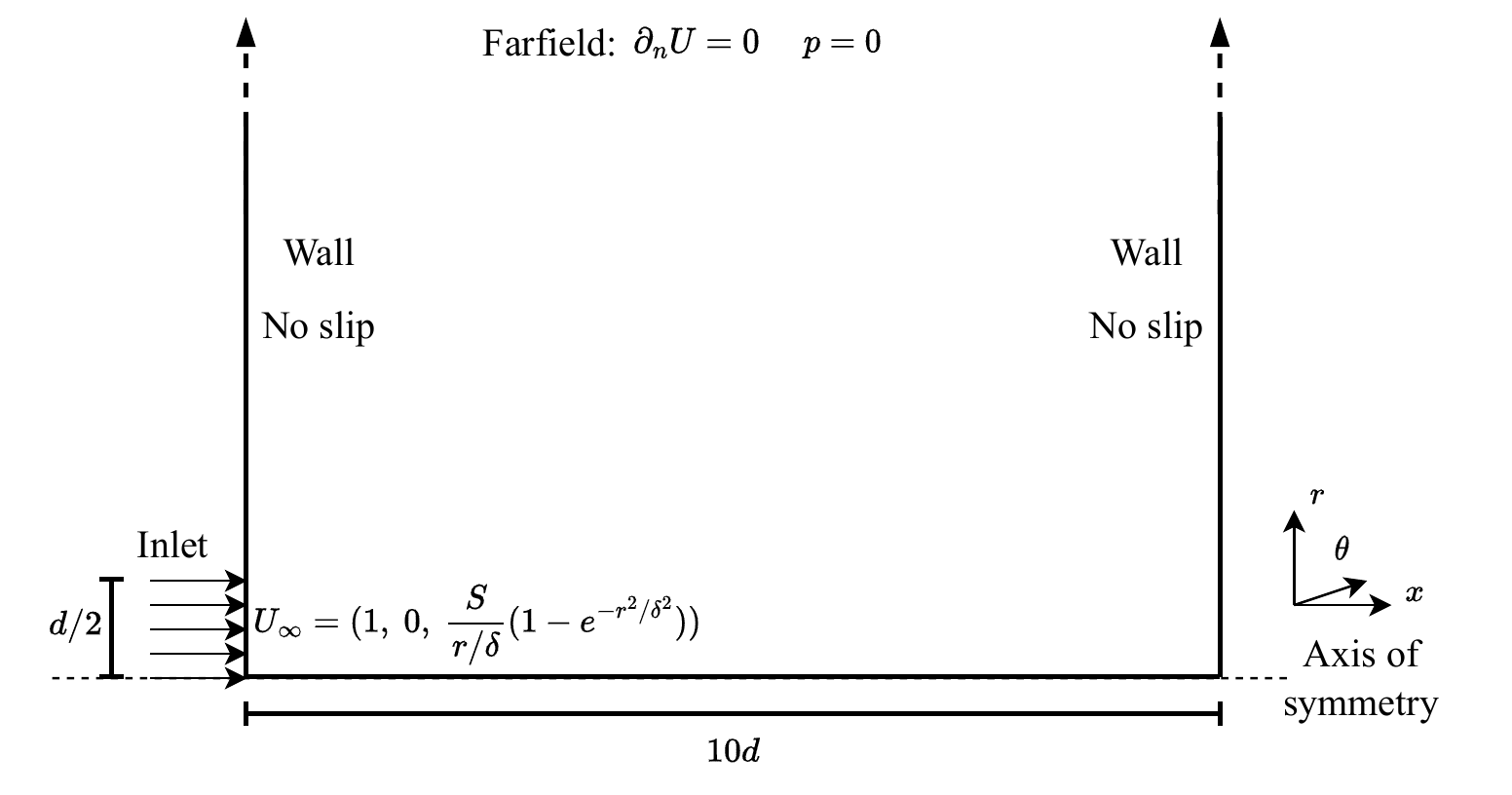}
    \caption{2D axisymmetric domain of a jet impinging on a wall.}
    \label{fig:jet-domain}
\end{figure}
The second flow considered is an axisymmetric swirling jet impinging on a flat plate. This jet flow is selected due to its complicated periodic time signature and the effect of the third velocity component, which arises due to the swirl. The mesh setup and boundary conditions in figure \ref{fig:jet-domain}, mimic those published in \citet{herrada-jet}: a jet with uniform axial flow enters the domain from an inlet of diameter $d$ and Reynolds number $(Re=Ud/\nu)$ 204. The swirl comes from an imposed vortex characterized by two nondimensional parameters. These comprise the swirl parameter $(S)$, which is proportional to the circulation of the vortex, and a vortex core radius $(\delta)$. The particular selection of $S$ and $\delta$ are 3 and 0.25, respectively, to recover the physics seen by \citet{herrada-jet}. These parameters are selected due to the resulting complicated but periodic flow. 

\citet{herrada-jet} characterized the jet by a large recirculating region along the central axis, called a ``vortex breakdown (VB) bubble". There are also several axisymmetric vortex bubbles that originate and decay close to the wall over multiple intervals in each cycle. \citet{herrada-jet} noted the significant outward convection of the circulation due to radial flow from the jet impingement.

\begin{figure}
    \centering
    \includegraphics[scale=0.8]{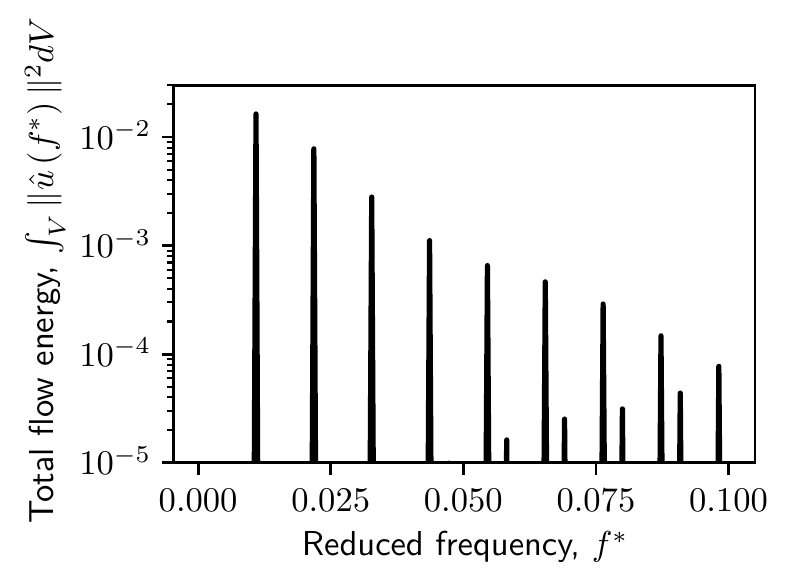}
    \caption{Energy contained in each Fourier mode of the swirling jet. Off-harmonic peaks seen between $0.05 < f^* < 0.1$ are due to aliasing but have negligible energy content.}
    \label{fig:jet_energy}
\end{figure}

The transient vortex bubbles result in a time signature that is composed of large energy concentrations at several discrete frequencies with a base frequency of $f^*=fd/U\approx 0.011$. The energy contained in each mode is shown in figure \ref{fig:jet_energy}. The modes decay exponentially in energy content with increasing frequency. The right bound of the plot represents the Nyquist frequency at $f^*=0.100$. Modes beyond this frequency are of extremely low magnitude and are thus they are ignored.

\begin{figure}
    \centering
	\begin{minipage}{0.33\textwidth}
		\centering
		\subcaptionbox{\hspace*{-1.75em}}{%
			\hspace{-0.25in}%
			\includegraphics[scale=0.8]{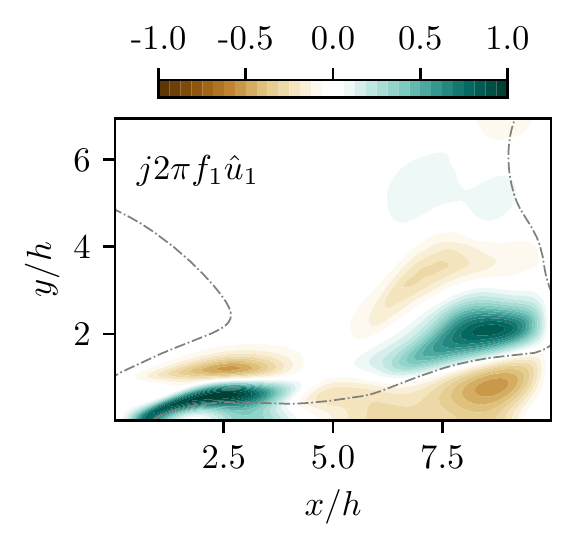}%
		}\qquad
	\end{minipage}\hfill
	\begin{minipage}{0.33\textwidth}
		\centering
		\subcaptionbox{\hspace*{-1.75em}}{%
			\hspace{-0.125in}%
			\includegraphics[scale=0.8,trim={0.3in 0 0 0},clip]{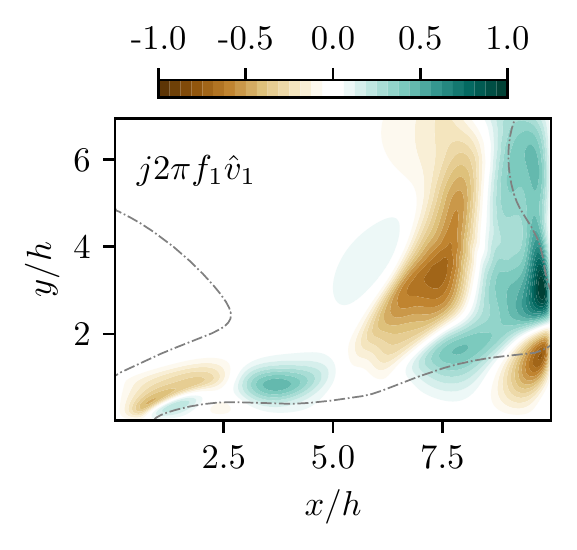}%
		}\qquad
	\end{minipage}
	\begin{minipage}{0.33\textwidth}
		\centering
		\subcaptionbox{\hspace*{-1.75em}}{%
			\hspace{-0.25in}%
			\includegraphics[scale=0.8,trim={0.3in 0 0 0},clip]{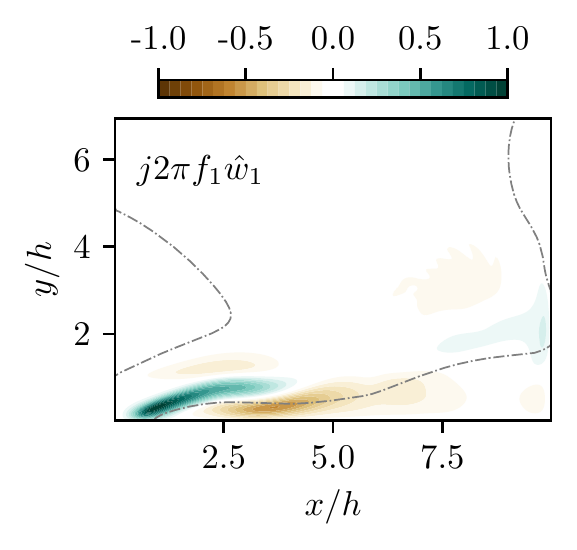}%
		}
	\end{minipage}
    \caption{Real value of FANS UTs of the swirling jet at the fundamental frequency in each coordinate direction: (a) axial, (b) radial, (c) azimuthal. Grey dashed line indicates $U = 0$ which is extent of VB bubble along the axis.}
    \label{fig:jet-UT-base}
\end{figure}
Figure \ref{fig:jet-UT-base} shows the values of the UTs in FANS at the fundamental frequency. These correspond to the highest energy peak shown in figure \ref{fig:jet_energy}. Significant axial velocity fluctuations are seen throughout the domain. Radial fluctuations are likewise present throughout but become stronger in magnitude downstream due to the presence of vortex bubbles close to the wall. Finally, the azimuthal fluctuations start strongly near the inlet and diminish toward the wall. This gradual diminishment of fluctuations is due to the outward convection of the swirl noted by \citet{herrada-jet}.

\begin{figure}
    \centering
	\begin{minipage}{0.5\textwidth}
		\centering
		\subcaptionbox{\hspace*{-1.75em}}{%
			\hspace{-0.25in}%
			\includegraphics[scale=0.8]{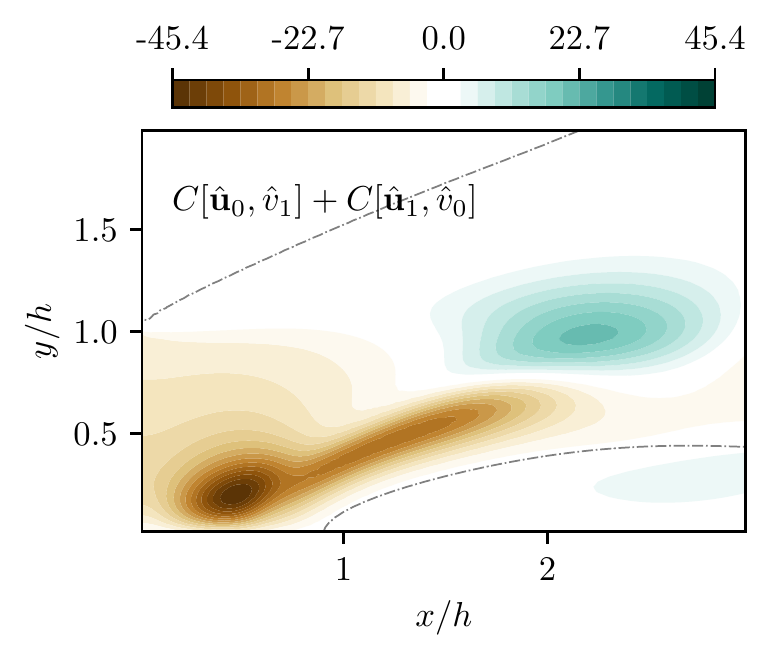}%
		}\qquad
	\end{minipage}\hfill
	\begin{minipage}{0.5\textwidth}
		\centering
		\subcaptionbox{\hspace*{-1.75em}}{%
			\hspace{-0.25in}%
			\includegraphics[scale=0.8]{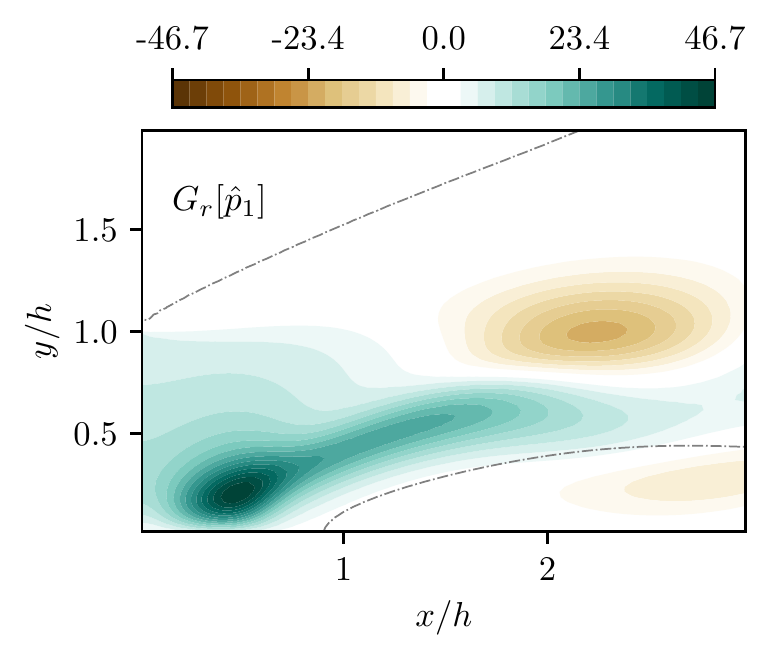}%
		}\qquad
	\end{minipage}
    \caption{Real value of FANS terms in the radial direction near the inlet of the swirling jet at the fundamental frequency: (a) Mean-flow convection, (b) pressure gradient. Grey line shows extent of VB bubble as in figure \ref{fig:jet-UT-base}.}
    \label{fig:jet-terms-radial}
\end{figure}
Figures \ref{fig:jet-terms-radial}(a) and (b) show the radial mean-flow convection and pressure gradient in the near-inlet region, respectively. These momentum fluxes are of considerably larger magnitude than the UT, peaking at over 45 times the maximum magnitude of the UT. The significant radial force fluctuations are due to the deflection of incoming flow against the VB bubble as indicated in figure \ref{fig:jet-terms-radial}. The interaction of the incoming jet with the VB bubble results in a large opposing pressure gradient. The relative magnitude of the forces combined with the phase of the UT (which can be compared through the sign of the value) shows that the pressure gradients dominate the convection in this region and drive the radial fluctuations. These radial fluctuations are important as they are related to the vortex bubbles that dominate the fluctuating flow. Thus, FANS can be used to elucidate the effect of large counteracting forces and compare them to the time dependent term, which represents the fluctuations at that particular timescale. 

\begin{figure}
    \centering
			\includegraphics[scale=1]{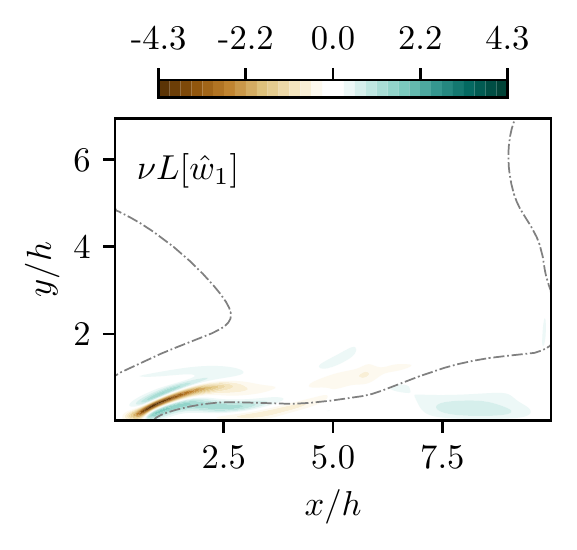}%
    \caption{Real value of FANS viscous diffusion term of the swirling jet in the azimuthal direction at the fundamental frequency.}    
    \label{fig:jet-terms-azimuthal}
\end{figure}

\citet{herrada-jet} found that this jet flow is highly sensitive to changes in viscosity. The significant viscous contribution to the momentum transfer close to the inlet in figure \ref{fig:jet-terms-azimuthal} implies a similar sensitivity. The phase and high magnitude of the azimuthal shear suggests that there is a dampening effect on the corresponding azimuthal fluctuations. This high shear stress shows the significant effect of the viscosity that results in high sensitivity of the flow physics to changes in the Reynolds number. In this way, FANS provides a platform to gain insight into the stability of the flow configuration.

The intermittent nature of the near-wall vortex bubbles observed in this flow results in a complicated time signature, where the vortex structures are observed across several frequencies. This suggests significant coupling between frequency components. To more thoroughly analyse this inter-frequency coupling of the jet flow and its relationship to the flow physics, Fourier stresses are analysed in more detail. Specifically, Fourier stresses represent the interactions that are generated by motion of the intermittent vortex bubbles. 

\begin{figure}
    \centering
	\begin{minipage}{0.33\textwidth}
		\centering
		\subcaptionbox{\hspace*{-1.75em}}{%
			\hspace{-0.25in}%
			\includegraphics[scale=0.8]{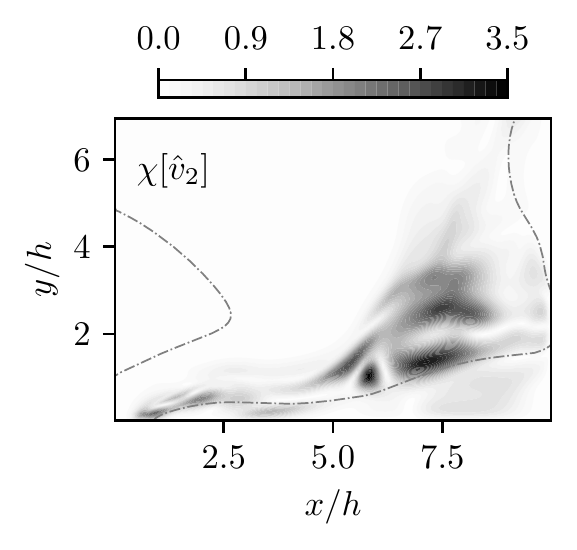}%
		}\qquad
	\end{minipage}\hfill
	\begin{minipage}{0.33\textwidth}
		\centering
		\subcaptionbox{\hspace*{-1.75em}}{%
			\hspace{-0.125in}%
			\includegraphics[scale=0.8,trim={0.3in 0 0 0},clip]{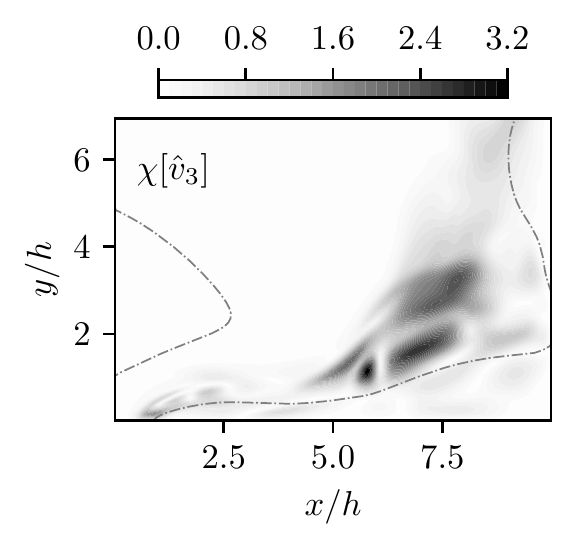}%
		}\qquad
	\end{minipage}
	\begin{minipage}{0.33\textwidth}
		\centering
		\subcaptionbox{\hspace*{-1.75em}}{%
			\hspace{-0.25in}%
			\includegraphics[scale=0.8,trim={0.3in 0 0 0},clip]{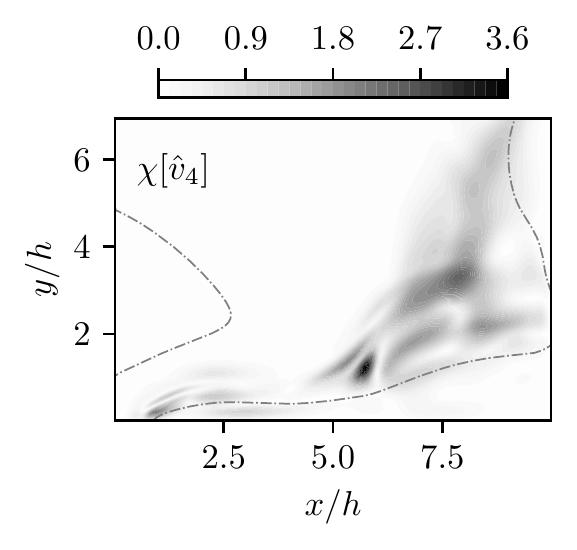}%
		}
	\end{minipage}
    \caption{Fourier stress magnitude in the radial direction of the swirling jet for (a) Mode 2, (b) Mode 3, (c) Mode 4.}
    \label{fig:jet-fstress-radial}
\end{figure}
The radial Fourier stress magnitude for modes 2-4 are shown in figure \ref{fig:jet-fstress-radial}. The radial terms are selected since they are characteristic of the vortex bubbles but not the swirling inlet flow. Notably, the vast majority of interactions occur outside of the recirculation region, which sits along the symmetry axis at $y/h=0$. This follows with the radial movement of the velocity fluctuations as seen in figure \ref{fig:jet-terms-radial}, which is due to interaction of the jet with the recirculation region and the wall. The Fourier stresses have significant magnitude at each mode. This is consistent with the expectation from a cascade of frequency content resulting in a large number of modes. Regions of elevated Fourier stresses are localized well downstream of the inlet and outside of the recirculation region. This matches the expected location of enhanced interaction due to fluctuations induced by the vortex bubbles. Thus, FANS analysis of the convective coupling coincides with physical intuition about the interactions and time signature of the flow.

\begin{figure}
    \centering
    \includegraphics[scale=0.8]{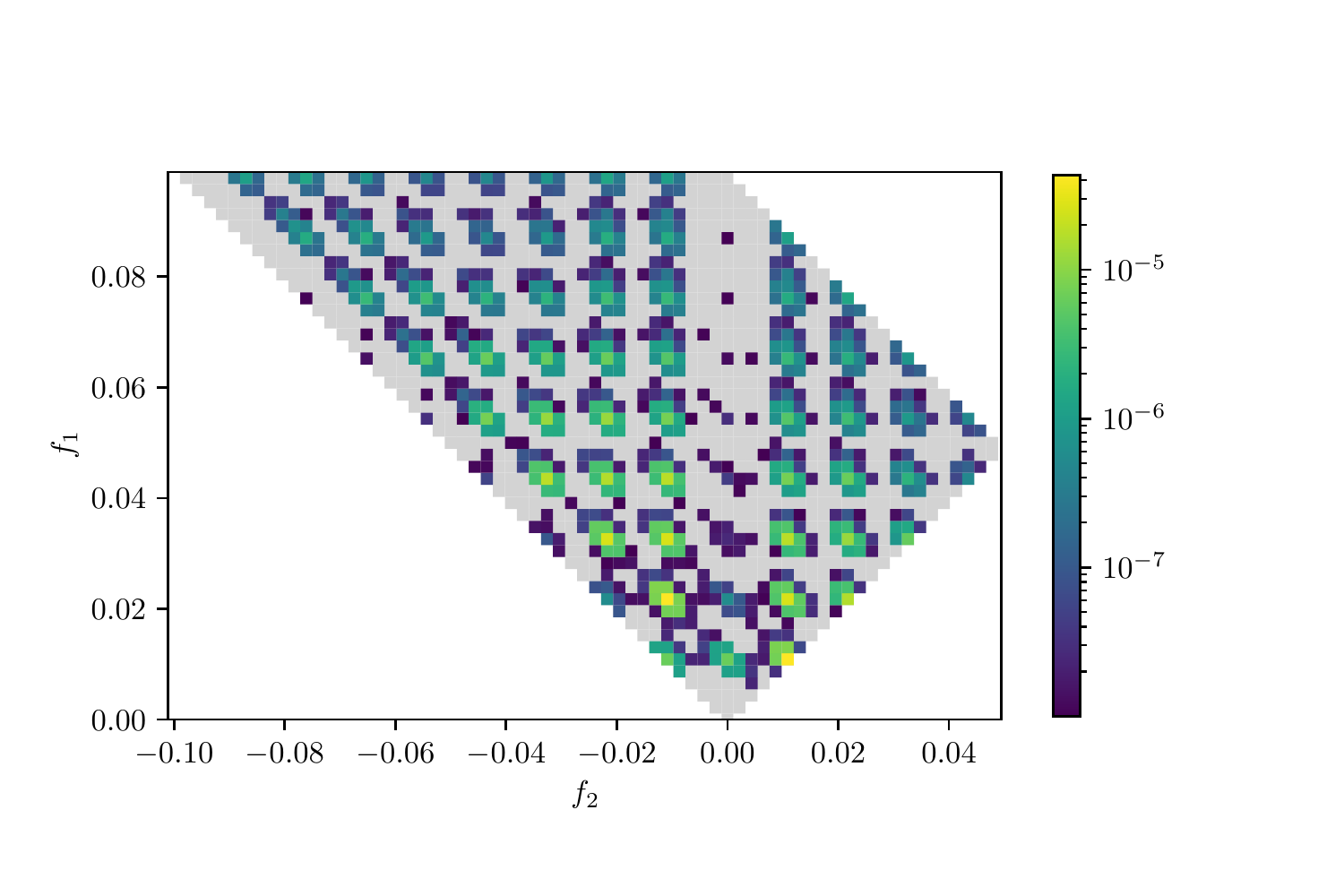}
    \caption{Mode bispectrum of the swirling jet flow.}
    \label{fig:jet-bmd}
\end{figure}

To further corroborate the influence of the vortex bubble on the inter-harmonic coupling, we explore the flow dynamics using BMD. Results of the bicorrelation coefficients $\lambda_1$ for the bispectrum analysis of the jet are shown in figure \ref{fig:jet-bmd}. As this flow is periodic, the BMD mode bispectrum shows a cascade of harmonics, starting with triads involving the fundamental frequency $f^*=fh/U\approx 0.011$. This agrees with the observations of interactions between several modes as seen in figure \ref{fig:jet-fstress-radial}.

\begin{figure}
    \centering
	\begin{minipage}{0.33\textwidth}
		\centering
		\subcaptionbox{\hspace*{-1.75em}}{%
			\hspace{-0.25in}%
			\includegraphics[scale=0.85]{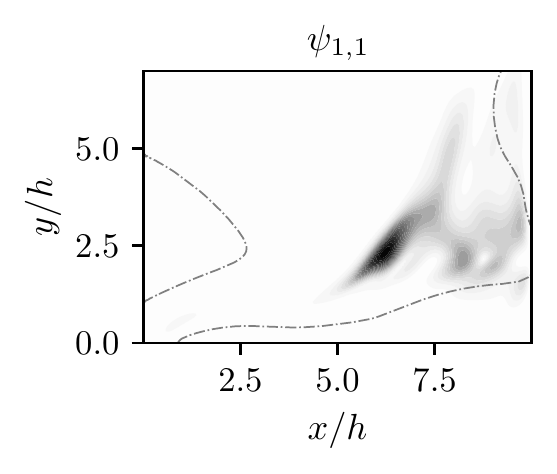}%
		}\qquad
	\end{minipage}\hfill
	\begin{minipage}{0.33\textwidth}
		\centering
		\subcaptionbox{\hspace*{-1.75em}}{%
			\hspace{-0.25in}%
			\includegraphics[scale=0.85]{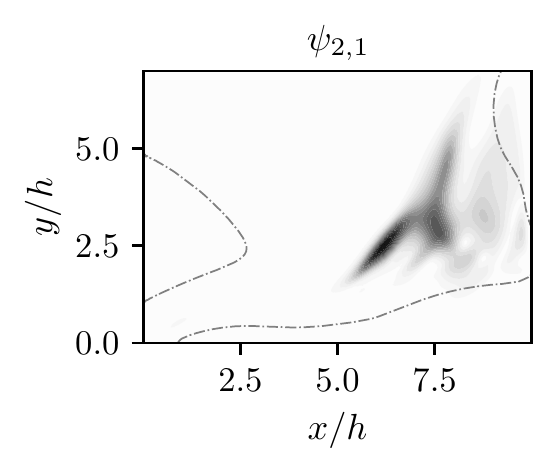}%
		}\qquad
	\end{minipage}
	\begin{minipage}{0.33\textwidth}
		\centering
		\subcaptionbox{\hspace*{-1.75em}}{%
			\hspace{-0.25in}%
			\includegraphics[scale=0.85]{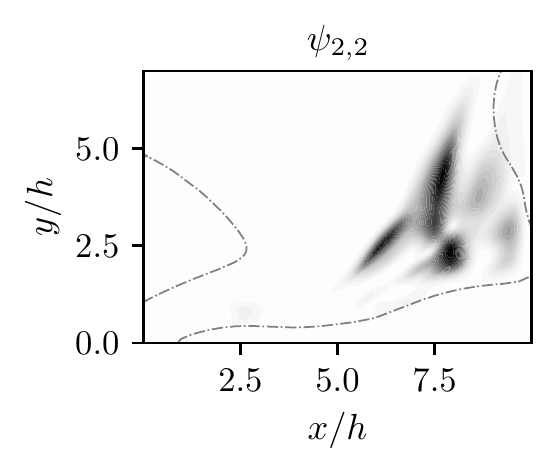}%
		}
	\end{minipage}
    \caption{Interaction maps in the radial direction of the swirling jet flow for triads (a)  $\psi_{1,1}$, (b) $\psi_{2,1}$ (c) $\psi_{2,2}$.}
    \label{fig:jet-imap-radial}
\end{figure}

Figure \ref{fig:jet-imap-radial} shows the BMD-calculated interaction maps for selected triads corresponding to $(f^*_1=f^*, f^*_2 = f^*, f^*_3 = 2f^*)$, $(f^*_1=2f^*, f^*_2 = f^*, f^*_3 = 3f^*)$, and $(f^*_1=2f^*, f^*_2 = 2f^*, f^*_3 = 4f^*)$. Figures \ref{fig:jet-imap-radial}(a-c) represent the radial component of the interaction map for each triad. Similar to FANS, the interactions are found to remain outside of the recirculation region. This is due to the outward forcing of the flow as discussed above. This finding about the interactions supports the conclusions brought by Fourier stresses. The BMD-calculated interaction maps also show the localization of interactions downstream ($x/h>5$) and outside of the recirculation region (figure \ref{fig:jet-imap-radial}). This suggests a similar relationship to the vortex bubbles to what was previously observed with radial Fourier stresses. The intermittent motion of the vortex bubbles results in significant interaction between modes, which then appears as enhanced triadic interactions or Fourier stresses detected by BMD and FANS, respectively.

\subsection{Side-by-side cylinders}
\begin{figure}
    \centering
    \includegraphics[scale=0.65]{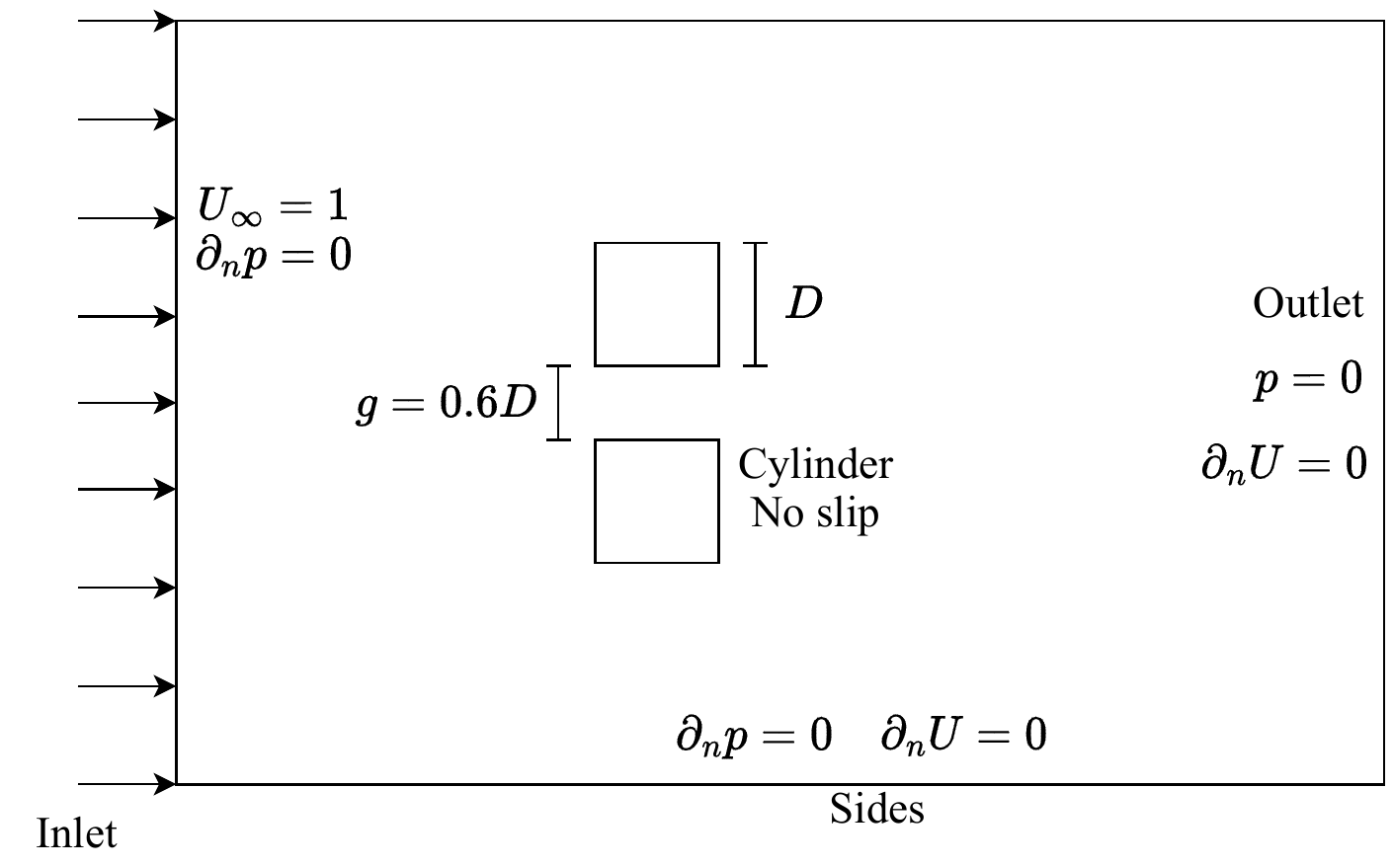}
    \caption{2D simulation domain of side-by-side cylinders with a gap ratio of 0.6.}
    \label{fig:pair-domain}
\end{figure}
The final case considered is that of two side-by-side cylinders of diameter $D$ at Reynolds number of $(Re=U_\infty D/\nu)$ 90. The cylinders are separated by a gap of $0.6D$. A representation of the computational domain is shown in figure \ref{fig:pair-domain}. These settings and the corresponding mesh requirements were taken from \citet{ma-pair}, against which the results are verified. This flow configuration results in an asymmetric, irregularly oscillating wake characterized by multiple shedding processes and their interaction. There is vortex shedding behind each of the two cylinders, which then merges into a single Karman street far downstream. There is also a ``flip-flopping" process where the wake asymmetry changes orientation \citep{ma-pair}. The flip-flop occurs randomly at a frequency much lower than that of the shedding \citep[see][]{burattini-flipflop,carini_flipflop}. As a result, there are multiple dominant frequencies. Depending on the location in the wake, the flow may be dominated by the low-frequency flip-flopping in between the cylinders, the high-frequency shedding process behind each of the cylinders, or the moderate-frequency shedding process in the wake far from the cylinders due to wake merging. This case is selected to investigate the momentum transport processes detailed by FANS in a nonperiodic case with multiple characteristic frequencies.

\begin{figure}
    \centering
	\begin{minipage}{0.5\textwidth}
		\centering
		\subcaptionbox{\hspace*{-1.75em}}{%
			\hspace{-0.125in}%
            \includegraphics[scale=0.8]{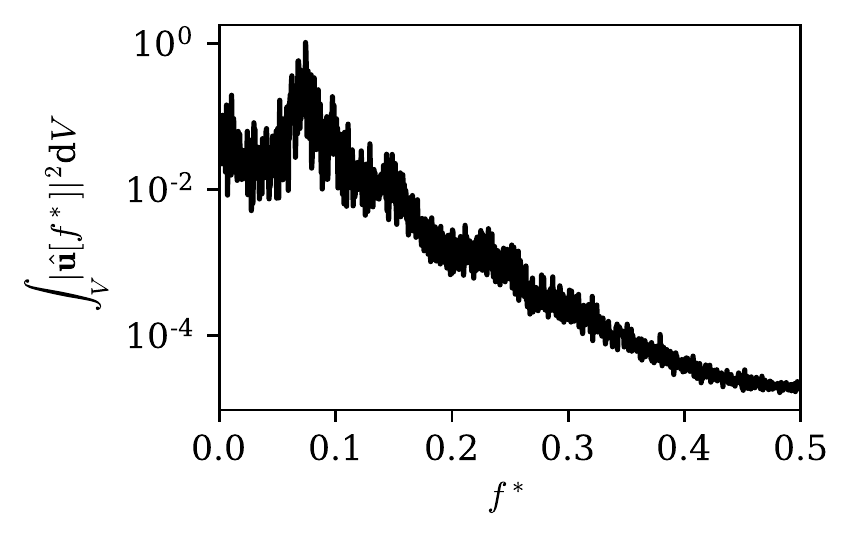}
		}\qquad
	\end{minipage}\hfill
	\begin{minipage}{0.5\textwidth}
		\centering
		\subcaptionbox{\hspace*{-1.75em}}{%
            \includegraphics[scale=0.8]{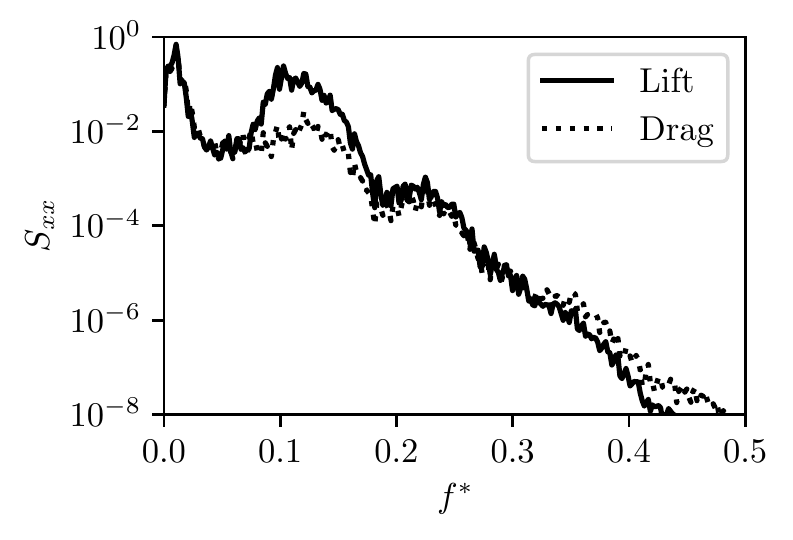}
		}\qquad
	\end{minipage}
    \caption{(a) Spectrum of flow energy fluctuations (b) power spectral density of drag and lift coefficients on the lower cylinder.}
    \label{fig:dual-psd}
\end{figure}
The spectrum of the flow energy fluctuations and the estimated power spectral density $(S_{xx})$ of the drag and lift coefficients on the lower of the two cylinders are shown in figures \ref{fig:dual-psd}(a) and (b) respectively. These broadband spectra show that the flow is irregular and not characterized by discrete frequency content. Within figure \ref{fig:dual-psd}(a), there is a broadband spectrum centred about $f^*=0.7$, whereas the drag and lift spectrum shows dominant frequencies that are centred around 0.01 and 1.2. The lowest frequency $(f^*=fD/U_{\infty}\approx 0.01)$ corresponds to the flip-flopping process, which occurs on irregular intervals at about this frequency. The moderate frequency shown in the energy spectrum corresponds to the moderate frequency shedding in the merged wake. The highest dominant frequencies in the lift and drag spectra ($0.9 < f^* < 1.5$) correspond to vortex shedding on the cylinder body. The flip-flopping and two vortex production processes will be analyzed using the FANS framework.

\begin{figure}
    \centering
    \includegraphics[scale=0.75]{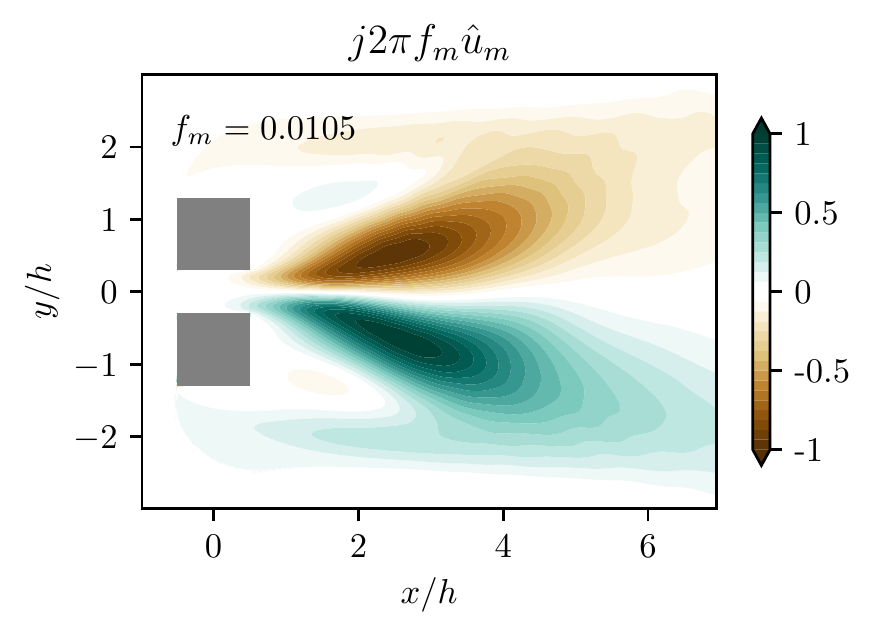}
    \caption{Real value of UT at flip-flopping frequency.}
    \label{fig:switch-UT}
\end{figure}
\begin{figure}
    \centering
	\begin{minipage}{0.5\textwidth}
		\centering
		\subcaptionbox{\hspace*{-1.75em}}{%
			\hspace{-0.125in}%
			\includegraphics[scale=0.75]{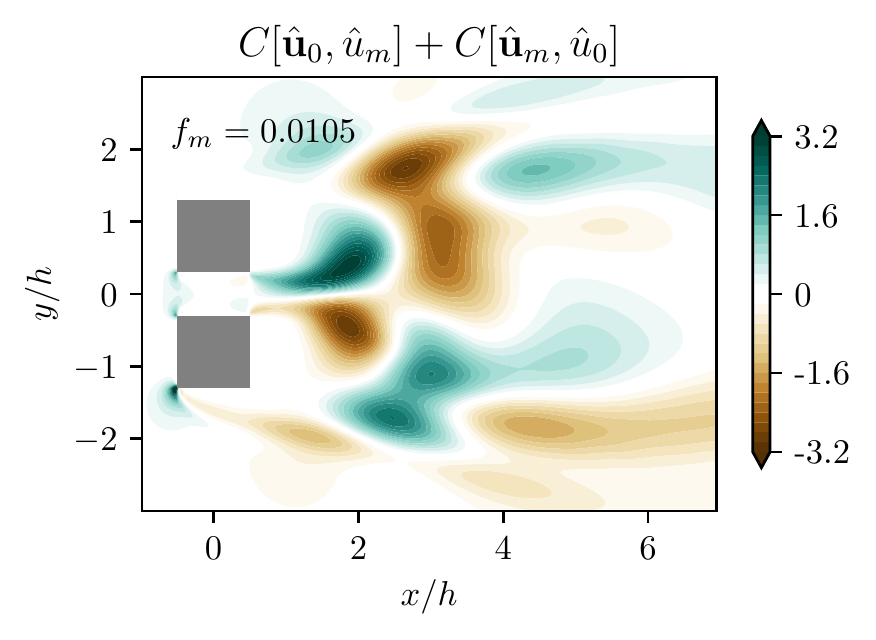}%
		}\qquad
	\end{minipage}\hfill
	\begin{minipage}{0.5\textwidth}
		\centering
		\subcaptionbox{\hspace*{-1.75em}}{%
			\includegraphics[scale=0.75,trim={0.7cm 0 0 0},clip]{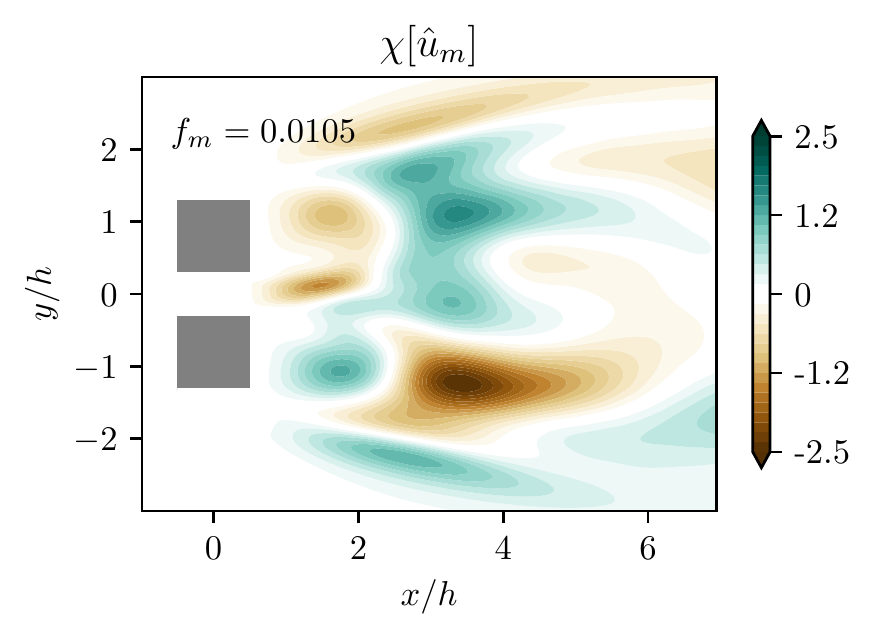}%
		}\qquad
	\end{minipage}\\
	\begin{minipage}{0.5\textwidth}
		\centering
		\subcaptionbox{\hspace*{-1.75em}}{%
			\includegraphics[scale=0.75]{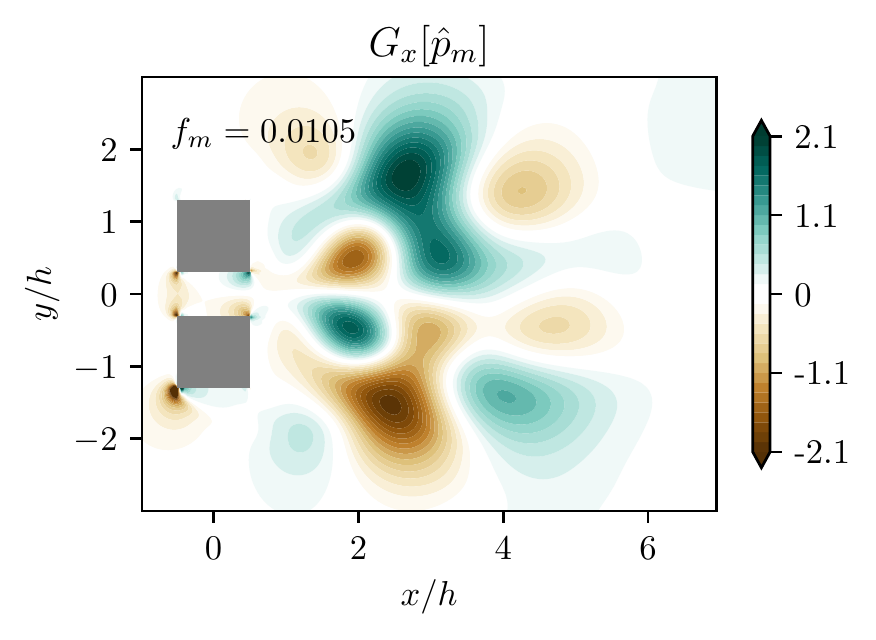}%
		}
	\end{minipage}
    \caption{Real value of FANS terms in the streamwise direction corresponding to flip-flopping frequencies: (a) Mean-flow convection, (b) Fourier stresses, (c) pressure gradient.}
    \label{fig:dual-terms-low}
\end{figure}
Figure \ref{fig:switch-UT} shows the UT at $f^*=0.0105$. This frequency corresponds to the flip-flopping process. Each of the lobes stemming from the gap represents a state of the flow, where the flow through the gap moves toward either the upper or lower cylinder at a given time. \citet{ma-pair} discusses the importance of the gap flow in this regime. Figures \ref{fig:dual-terms-low}(a-c) show the momentum fluxes in the streamwise direction at the flip-flopping frequency, where the mean-flow convection (figure \ref{fig:dual-terms-low}a), Fourier stresses (figure \ref{fig:dual-terms-low}b) and pressure gradient (figure \ref{fig:dual-terms-low}c) are significant. The diffusion does not contribute significantly to the transport process. 
These momentum fluxes show the importance of the gap flow to the flip-flopping effect. Large velocity gradients on the leeward side of the gap result in the large convective flux seen in figure \ref{fig:dual-terms-low}(a) that sustains the switching process. These large velocity gradients are due to the combination of significant gap flow and tight spacing between the cylinders. This combination was highlighted by \citet{ma-pair} as important to the instability that leads to the irregular flow, which is reflected here.

There are also significant Fourier stresses at this frequency, immediately behind the cylinders. These stresses represent interactions of different frequencies within the bands of shedding and Karman street frequencies. Since the bandwidth of the energy and force coefficient spectra is larger than the flip-flopping frequency, interactions between modes that lie within these bands result in Fourier stresses at this frequency.

\begin{figure}
    \centering
	\begin{minipage}{0.5\textwidth}
		\centering
		\subcaptionbox{\hspace*{-1.75em}}{%
			\hspace{-0.125in}%
			\includegraphics[scale=0.75]{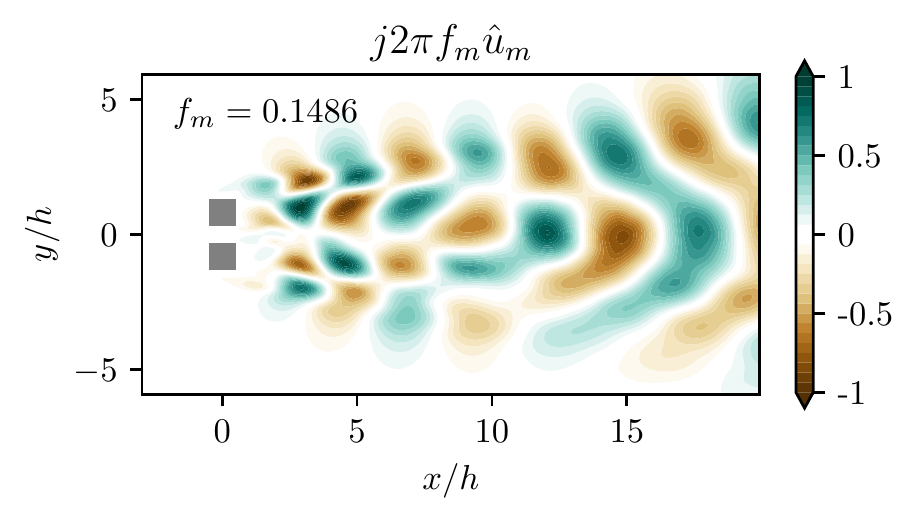}%
		}\qquad
	\end{minipage}\hfill
	\begin{minipage}{0.5\textwidth}
		\centering
		\subcaptionbox{\hspace*{-1.75em}}{%
			\includegraphics[scale=0.75,trim={0.7cm 0 0 0},clip]{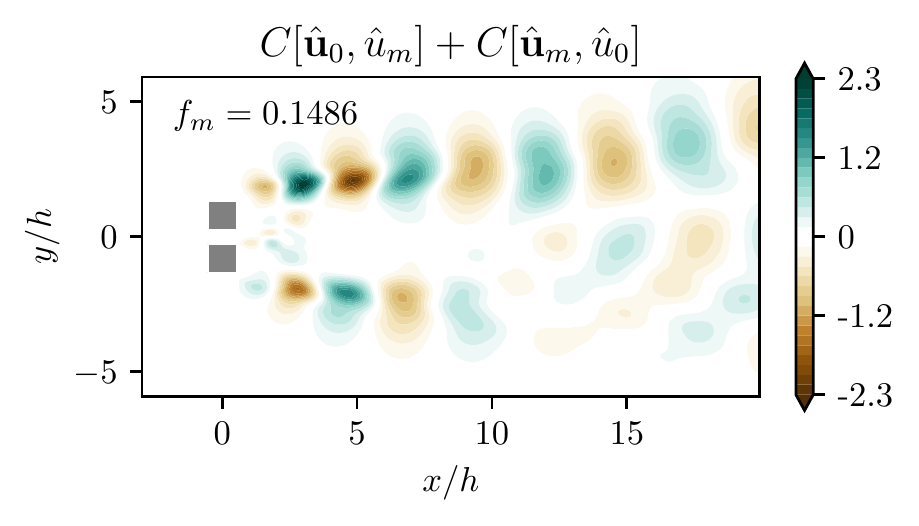}%
		}\qquad
	\end{minipage}\\
 	\begin{minipage}{0.5\textwidth}
		\centering
		\subcaptionbox{\hspace*{-1.75em}}{%
			\hspace{-0.125in}%
			\includegraphics[scale=0.75]{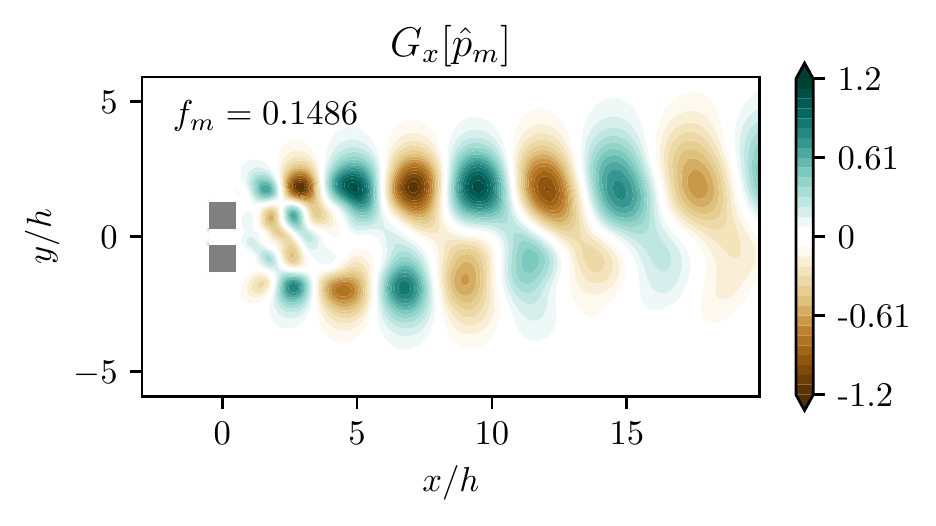}%
		}\qquad
	\end{minipage}\hfill
	\begin{minipage}{0.5\textwidth}
		\centering
		\subcaptionbox{\hspace*{-1.75em}}{%
			\includegraphics[scale=0.75,trim={0.7cm 0 0 0},clip]{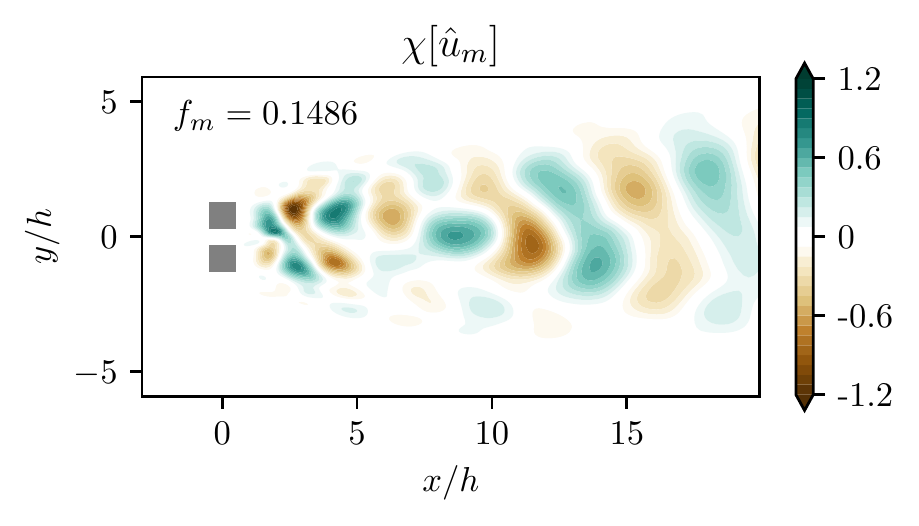}%
		}\qquad
	\end{minipage}
    \caption{Real values of streamwise momentum fluxes at $f^*_m=0.148$: (a) UT, (b) mean-flow convection, (c) pressure, (d) Fourier stresses.}
    \label{fig:dual-terms-high}
\end{figure}
Figure \ref{fig:dual-terms-high} shows the significant momentum fluxes at $f^*=0.148$. This frequency corresponds to vortex shedding from the individual cylinders, which appears in the force spectra in figure \ref{fig:dual-psd}(b). Evidence of shedding near the cylinders can be seen in the UT and pressure gradient in figure \ref{fig:dual-terms-high}. This aligns with our expectations due to the force fluctuations at this frequency.
This mode persists into the far wake, where the Karman vortex street is dominant (figure \ref{fig:dual-terms-high}(a). This persistence shows that this mode has an effect on the Karman vortex street. This effect is not immediately apparent from the energy spectrum in figure \ref{fig:dual-psd}(a) due to dominance of the lower frequency shedding at $f^*\approx0.73$ on the overall velocity fluctuations in the wake.

\begin{figure}
    \centering
	\begin{minipage}{0.5\textwidth}
		\centering
		\subcaptionbox{\hspace*{-1.75em}}{%
			\hspace{-0.125in}%
			\includegraphics[scale=0.75]{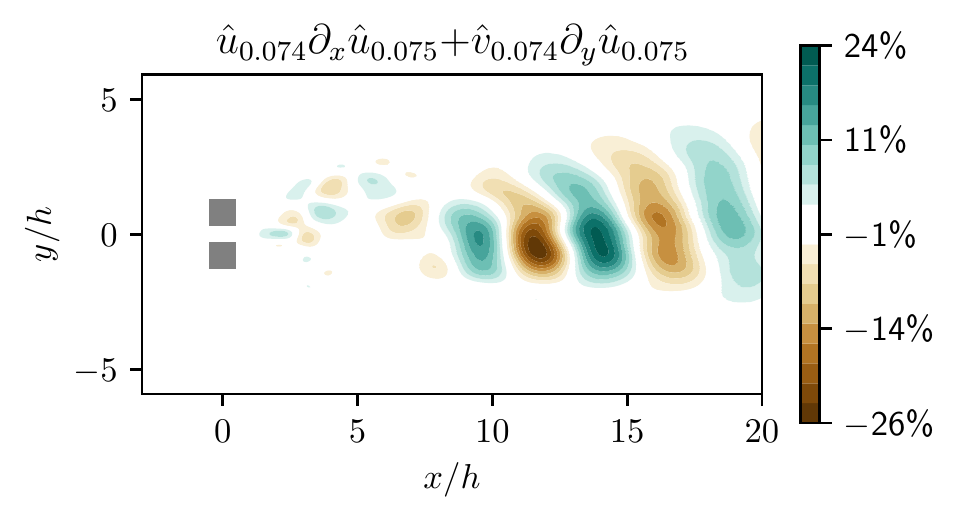}%
		}\qquad
	\end{minipage}\hfill
	\begin{minipage}{0.5\textwidth}
		\centering
		\subcaptionbox{\hspace*{-1.75em}}{%
			\includegraphics[scale=0.75,trim={0.7cm 0 0 0},clip]{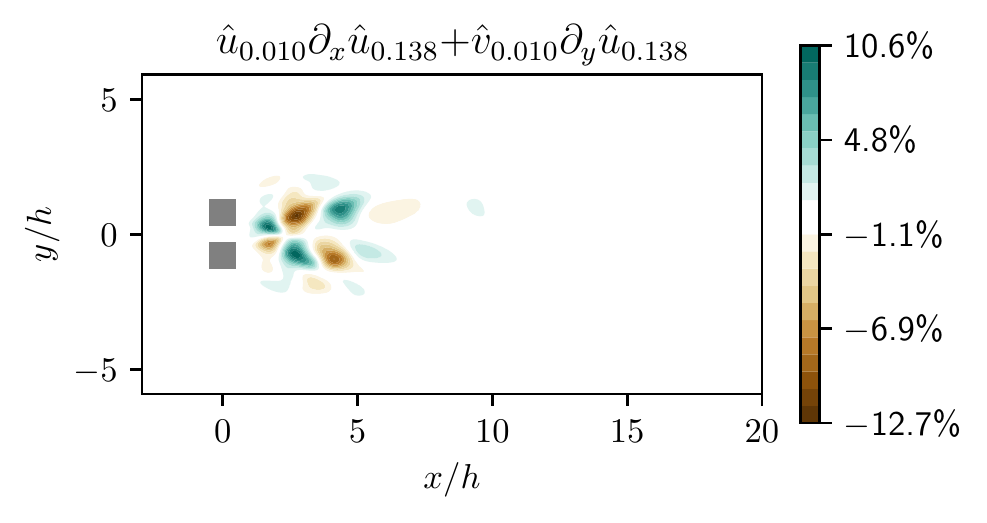}%
		}\qquad
	\end{minipage}
    \caption{Real values of components of Fourier stresses at $f^*=0.148$ in the streamwise direction (a) $(f^*_1=0.074, f^*_2=0.075)$ (b) $(f^*_1=0.011, f^*_2=0.138)$.}
    \label{fig:dual-fstress-components}
\end{figure}

The Fourier stresses in figure \ref{fig:dual-terms-high}(d) extend throughout the domain and have similar magnitude to the UT. The localization of wake phenomena, such as flip-flopping and the Karman street, suggest that the Fourier stresses at this frequency are due to a broad range of interactions. FANS provides a method to detect the primary mechanisms that contribute to these stresses in the form of individual terms in the Fourier stress term. Select terms are shown in figures \ref{fig:dual-fstress-components}. Modes are subscripted with their frequency due to the broadband spectrum of this flow. For instance, $\hat{u}_{0.075}$ is the streamwise direction of the velocity Fourier mode at $f^*_m=0.075$. 
The Fourier stress terms corresponding to the contribution of the main Karman shedding frequency in the far wake is represented in figure \ref{fig:dual-fstress-components}(a). The momentum flux due to this term shows increased magnitude in the region of $x/h>10$. The increased stress magnitude in this region is related to the formation of the Karman street and the interaction between resultant vortices. Thus, it may be said that the primary driver of Fourier stresses in the far wake is the existence of Karman vortices in this region.

Figure \ref{fig:dual-fstress-components}(b) shows the momentum flux in the same Fourier stress term due to a triad, $(f^*_1=0.010, f^*_2=0.138, f^*_3=0.148)$. The low frequency of $0.010$ corresponds to the flip-flopping process. The higher frequency $0.138$ is within the frequency band of lift and drag fluctuations. This triad represents one of several similar interactions between these processes (other relevant triads are not shown here for brevity). Magnitude of the convective flux shown in figure \ref{fig:dual-fstress-components}(b) shows that the interaction between these modes is a significant source of momentum for the fluctuations at $f^*\approx0.148$ immediately behind the cylinders. This suggests that the gap flow and flip-flopping process modulates the vortex shedding. This is supportive of the findings of \citet{ma-pair}, who also reported this modulation of the shedding behind the cylinders. 

\begin{figure}
    \centering
    \includegraphics[scale=0.8]{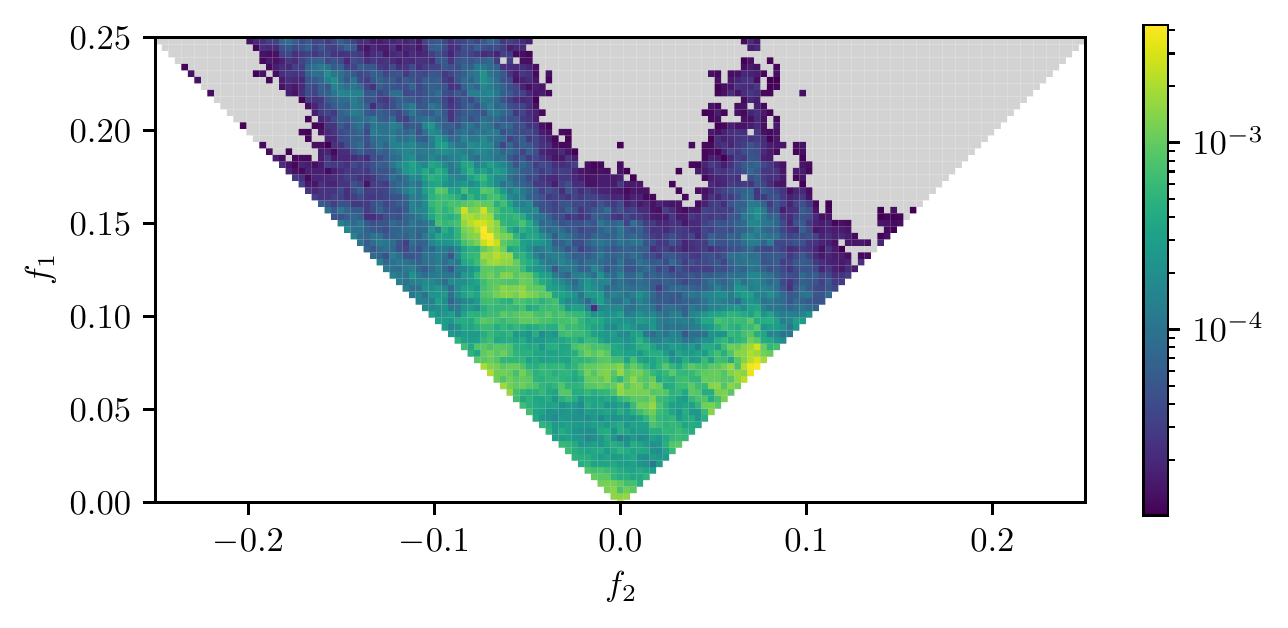}
    \caption{BMD mode bispectrum for the dual cylinder case. Spectrum is blanked below $10^{-5}$ to highlight strongly interacting triads.}
    \label{fig:dual-BMD}
\end{figure}

We once again utilize BMD to explore the flow dynamics. The spectrum of $\lambda_1$ is shown in figure \ref{fig:dual-BMD}.  There are local maxima in the bispectrum involving frequencies at $f^*\approx 0.73$. The diagonal and vertical bands of elevated correlation around these maxima may be due to spectral leakage \citep{schmidt-bmd}. Locations of the maxima correspond to triads around $(f^*_1\approx 0.73, f^*_2 \approx 0.73, f^*_3 \approx 0.147)$ and $(f^*_1\approx 0.147, f^*_2 \approx -0.73, f^*_3 \approx 0.73)$. These triads involve the same frequencies, indicating that they are mirrored and capture the same interactions. 

\begin{figure}
    \centering
	\begin{minipage}{0.5\textwidth}
		\centering
		\subcaptionbox{\hspace*{-1.75em}}{%
			\hspace{-0.125in}%
			\includegraphics[scale=0.75]{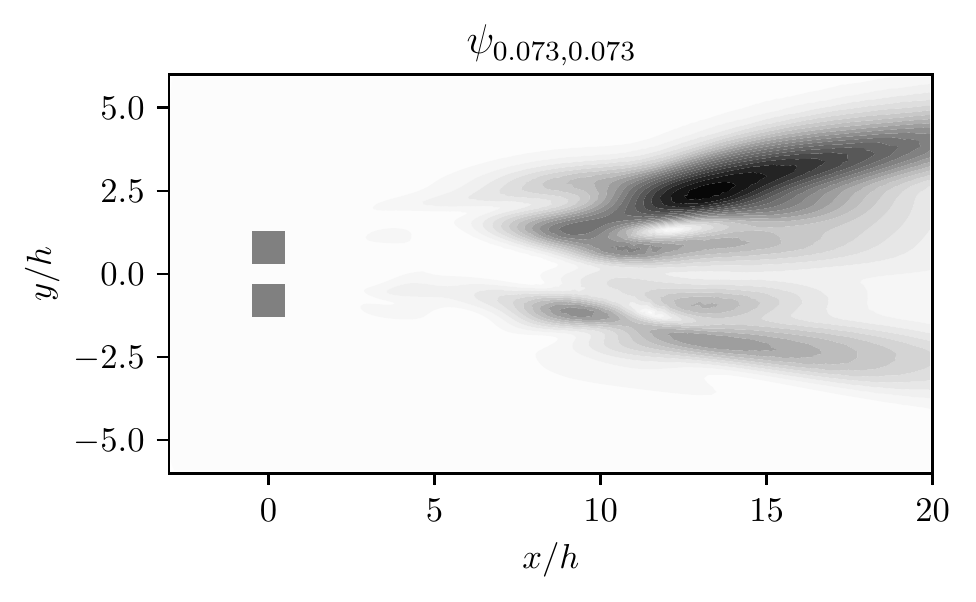}%
		}\qquad
	\end{minipage}\hfill
	\begin{minipage}{0.5\textwidth}
		\centering
		\subcaptionbox{\hspace*{-1.75em}}{%
			\includegraphics[scale=0.75,trim={0.7cm 0 0 0},clip]{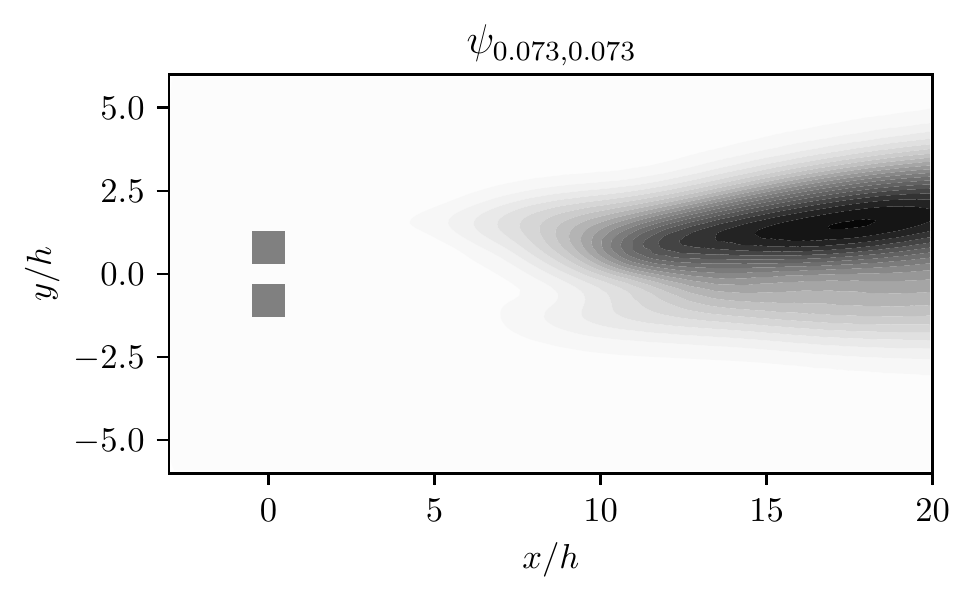}%
		}\qquad
	\end{minipage}
    \caption{Bispectral interaction maps corresponding to far-wake frequencies, represented by the triad $(f^*_1=0.073, f^*_2=0.073, f^*_3=0.147)$: (a) streamwise (b) transverse components.}
    \label{fig:bmd-imap-11}
\end{figure}

Interaction maps corresponding to the triad $(f^*_1\approx 0.73, f^*_2 \approx 0.73, f^*_3 \approx 0.147)$ are shown in figure \ref{fig:bmd-imap-11}. These correspond to the same wake processes discussed above using FANS. The exact listed frequency is slightly different from the FANS analysis due to the change in resolution, where BMD has lost some frequency resolution due to utilization of multiple windows. The interaction maps show significant triadic interaction beyond $x/D\approx 10$. This corresponds to the merger region, where the wakes of the individual cylinders combine and turn into a single Karman street. The driving frequency of $f^*_1=0.73$ is characteristic of the vortex street. This frequency is outside of the range of important force spectra seen in figure \ref{fig:dual-psd}. The resultant frequency of $f^*_3=0.147$ is characteristic of the cylinder-shedding, which continues to have an influence downstream, as was deduced from the FANS analysis. The maps show interaction between these modes in the vortex street far from the cylinders. This interaction is consistent with the behaviour of the Fourier stresses in figure \ref{fig:dual-fstress-components}(a).

\section{Conclusions}
Periodic and turbulent flows exhibit complicated interactions that can be difficult to characterize. Here, the Fourier-Averaged Navier Stokes (FANS) equations are introduced to provide insight into the physics of flows with cyclic characteristics, including both regular (periodic) and irregular (non-periodic) recurrences. By combining information from Fourier modes and the Navier-Stokes equations, this method produces momentum equations for individual Fourier modes. The equations contain pressure, diffusion, and convection terms that can be related to the velocity fluctuations and nonlinear interactions between modes. By treating the equations as a momentum budget, it is possible to directly compare different forces that contribute to fluctuations at a particular timescale. Analysing these FANS terms is shown to be a strong method with easier application and interpretability. The method directly addresses nonlinear interactions due to convection and relate them to the magnitude of other forces. FANS also includes phase information that can indicate when forces may be counteracting. This eases the physical interpretation of flow physics by isolating separate timescales and directly relating mode shapes to the governing equations. This is illustrated for periodic flows through case studies on the wake of a square cylinder and a swirling jet. FANS analysis can also be utilized for irregularly oscillating flows, which is shown through analysis of nonperiodic flow around two side-by-side cylinders. The method is shown to replicate results of other methods in all three case studies with minimal effort due to its simple calculation, and connections to the governing equations of fluid mechanics. Using the FANS equations is a viable method to identify and describe the relationships between different processes in flows with regularly or irregularly repeating characteristics.

\section*{Funding}
This work was funded by the Natural Sciences and Engineering Research Council of Canada.

\section*{Declaration of interest}
The authors report no conflict of interest.

\bibliographystyle{abbrvnat}
\bibliography{jfm}
\end{document}